\documentclass[epjCONF]{svjour}
\usepackage{graphics}
\usepackage[varg]{txfonts} 
\usepackage[latin1]{inputenc}
\session-title{UHECR 2012}
\begin{document}
\title{Measurement of Ultra-High Energy Cosmic Rays:\\
An Experimental Summary and Prospects}
\author{M. Fukushima
\inst{1}
\fnmsep\thanks
{\email{fukushim@icrr.u-tokyo.ac.jp}} }
\institute{Institute for Cosmic Ray Research, University of Tokyo,\\
Kashiwanoha 5-1-5,
Kashiwa, Chiba, 277-8582 Japan}
\abstract{ 
Measurements of Ultra-High Energy Cosmic Rays  achieved 
remarkable progress in the last 10 years. 
Physicists, gathered from around the world in the symposium UHECR-2012 held
at CERN on February 13-16 2012, 
reported their most up-to-date observations, discussed the meaning
of their findings, and identified remaining problems and
future challenges in this field. This paper is a part of the symposium 
proceedings on the experimental summary and future prospects
of the UHECR study.
}
\maketitle

\section{Introduction}
\label{introduction}
Cosmic Rays (CRs) are high energy radiations, mostly
made up of protons and nuclei, arriving at the Earth
isotropically from  outer space. Their energies range
from 10$^{9}$ eV to 10$^{21}$ eV, with a steeply
and monotonously decreasing flux approximately proportional
to E$^{-3}$. Ultra-High Energy Cosmic Rays (UHECRs)
reaching an energy of 10$^{20}$ eV start interacting
with the Cosmic Microwave Background (CMB) 
during their propagation in extra-galactic space,
produce nuclear $\Delta$ resonance and lose part
of their energy in the decay of the $\Delta$. 
In 1966, just after the discovery of
the CMB, Greisen, Zatsepin and Kuzmine (GZK)
predicted that this interaction causes a strong
suppression of the CR flux at around 10$^{20}$ eV
(GZK cutoff) \cite{gzk}. 
The search for this ``end-point'' in the CR spectrum
has been one of the most important subjects of
cosmic ray physics.

Cosmic Rays were discovered by V.~Hess 100 years ago in 1912
using an electroscope on board a balloon
reaching 5,400 m altitude.  Extensive Air Showers (EAS), particle
cascades caused by the high energy primary CRs interacting
with an atmospheric nucleus, were identified by P.~Auger
in 1938 using a coincidence of Geiger counters separated by 
up to 150 m on the ground \cite{auger}. 
The existence of UHECRs reaching
10$^{20}$ eV was first reported by J.~Linsley in 1962,
using a network of scintillation counters deployed
over an area of 8 km$^2$ \cite{linsley}.

\section{Ultra-High Energy Cosmic Rays}
\label{uhecr}
UHECRs are extremely rare; they are detected
as  Extensive Air Showers reaching the ground
over 100 km$^2$ area approximately once per year. 
A traditional way of detecting UHECRs is by using an array of Surface Detectors (SDs) sparsely deployed on the ground. The energy and the arrival direction of
the primary CR are reconstructed from the number of
charged particles and their arrival times measured by the SDs.
One of the largest arrays using this
technique, AGASA with 100 km$^2$ covered area using
optical fiber network, reported a possible extension of the
UHECR energy spectrum beyond the predicted GZK cutoff
in 1998 \cite{agasa}.  

A different method of detecting EAS was discussed by
A.~Chudakov, M.~Oda and K.~Suga in 1958-1962;
they proposed to detect scintillation (fluorescence) light
produced by the passage of
EAS particles in the atmosphere by using an imaging
telescope \cite{norikura}.
The first detection of EAS by air fluorescence was
made in Japan by G.~Tanahashi in 1968 \cite{tanahashi},
who came back from Cornell University where an intensive study
of air fluorescence detection had been initiated by
K.~Greisen. The method was systematically
developed and refined by the Fly's Eye group in Utah, and
the first observation of the GZK cutoff was announced in 2008
\cite{hiresgzk} 
by the upgraded detector, High Resolution Fly's Eye (HiRes).

The history of EAS observation and fluorescence
method was reviewed by {\it P.~Sokolsky}
\footnote{The name in italic is meant his/her contribution at
the UHECR-2012 is available in the printed proceedings 
or in the form of presentation file at the symposium website.}
at the beginning of UHECR-2012
(for a historical review of EAS,
see also \cite{kampert-watson} by K.~H.~Kampert
and A.~A.~Watson). 
Recent efforts of standardizing
the air fluorescence yield and its dependence on the atmospheric
condition were reported by {\it B.~Keilhauer}. 

The emerging air
fluorescence method enabled:  
\begin{itemize}
\item
{\bf experimental determination of the primary CR energy}
by measuring the total amount of fluorescence light in the 
atmosphere (total absorption calorimetry), and
\item
{\bf statistical determination of the primary particle species}, 
or the mass composition, by measuring the longitudinal shower
development in the atmosphere (X$_{max}$ measurement, see below).
\end{itemize}

Modern large-scale UHECR experiments, 
Pierre Auger Observatory (hereafter called Auger) and
Telescope Array experiment (TA), are hybrid experiments
employing a large SD array and a battery of Fluorescence Detectors
(FDs) overlooking the SD area. 
The SD array has a high duty factor ($\sim$100\%) and has
flat acceptance above a certain threshold energy. The FD
is operated only on clear, moonless nights (duty $<$ 10\%), and
its acceptance grows with the energy of CRs.
The number of SD/FD coincident events (hybrid events) are
limited, but they are important as they carry
the largest amount of information on the produced EAS.
For Auger and TA, high statistics analyses of energy
spectrum and anisotropy are performed using the SD data,
with its energy calibrated by the FD data using hybrid events. 
The mass composition is determined by using
average X$_{max}$, $<X_{max}>$, and its distribution width, RMS(X$_{max}$),
observed by the FD, where X$_{max}$ [g/cm$^2$] is defined as the slant
atmospheric depth of the maximum shower development
in the atmosphere.

The SD array of Auger is composed of 1600 water Cherenkov counters
deployed in a triangular grid of 1.5 km spacing, and
covers a ground area of 3000 km$^2$. 
Each tank contains water in a bag with 10 m$^2$ bottom area
and 1.2 m height. The SD of TA
is composed of 507 scintillation counters deployed in a mesh with 1.2 km
spacing, and covers an area of 700 km$^2$. Each TA/SD has two layers
of plastic scintillators, 1.2 cm thick and 3 m$^2$ large.
Water Cherenkov counters and plastic scintillators are sensitive to both the
Electro-Magnetic (EM) component (electrons and gammas) and hadronic component
(muons) in the EAS.  The Cherenkov signal in the water tank is larger
for penetrating hard muons than soft electrons and gammas in the shower,
thus making the Auger/SD more sensitive to the hadronic component in the EAS.
The plastic scintillator of TA equally samples the charged particles,
and the resultant signal from the EAS is dominated by the 
outnumbered electrons and photons.

The number of observed UHECRs increased by more than an order of magnitude
by the new experiments, Auger and TA, and the quality of UHECR observation
improved largely by the simultaneous use of the ground array (SD) and air
fluorescence telescope (FD). A combination of the 
two experiments, Auger in Argentina
(35.3$^\circ$ South) and TA in Utah, USA (39.4$^\circ$ North)
now covers the entire sky for CR observations.

\section{Symposium UHECR-2012}
\label{symposium}
The symposium UHECR-2012 was held at CERN on
February 13-16 2012
\footnote
{http://indico.cern.ch/conferenceDisplay.py?confId=152124
}
as a follow up of UHECR-2010 held in Nagoya in December 2010
\footnote
{http://uhecr2010.icrr.u-tokyo.ac.jp/ and 
http://proceedings.aip.org/resource/2/apcpcs/1367/1?
}.

There were 230 participants from 25 countries, 
and 53 oral and 48 poster presentations were made.
Five working groups (WGs) were formed before the symposium
from members of Auger, HiRes, TA,
Yakutsk air shower array
(Yakutsk) and the very forward LHC experiment (LHCf).
The physics subjects of the WGs
are listed in Table-\ref{tab:5wgs}, together with
WG members and convenors.
\begin{table}
\caption{UHECR-2012 Working Groups.}
\label{tab:5wgs}
\begin{tabular}{lll}
\hline
\noalign{\smallskip}
Subject & Convenors & Members \\
\noalign{\smallskip}\hline\noalign{\smallskip}
Spectrum & B.~R.~Dawson$^1$, &
   T.~Abu-Zayyad$^{2,3}$, D.~Ikeda$^3$, D.~Ivanov$^{2,3}$, I.~C.~Maris$^1$, \\
         & Y.~Tsunesada$^3$ &
   M.~Pravdin$^4$, M.~Roth$^1$, A.~Sabourov$^4$, F.~Salamida$^1$ \\
\noalign{\smallskip}\hline\noalign{\smallskip}
Composition & J.~Bellido$^1$,   &
   E.~Barcikowski$^3$, Y.~Egorov$^4$, S.~Knurenko$^4$, V.~de~Souza$^1$, \\
            & J.~W.~Belz$^{2,3}$ &
   Y.~Tameda$^3$, Y.~Tsunesada$^3$, M.~Unger$^1$ \\
\noalign{\smallskip}\hline\noalign{\smallskip}
Anisotropy & P.~Sommers$^1$,  &
   O.~Deligny$^1$, A.~Ivanov$^4$, J.~de~Mello~Neto$^1$, \\
           & P.~Tinyakov$^3$, &
   H.~Sagawa$^3$, L.~Timofeev$^4$, I.~Tkachev$^3$ \\
\noalign{\smallskip}\hline\noalign{\smallskip}
Multi-Messenger & M.~Risse$^1$, &
   J.~Alvarez-Muniz$^1$, B.~T.~Stokes$^{2,3}$ \\
                & G.~I.~Rubtsov$^{3,4}$ & \\
\noalign{\smallskip}\hline\noalign{\smallskip}
Shower Simulation and & R.~Engel$^1$ &
   J.~Allen$^1$, A.~Castellina$^1$, K.~Itakura$^{\rm *}$, K.~Kasahara$^{3,5}$, \\
Hadronic Interactions &              &
   S.~Knurenko$^4$, S.~Ostapchenko$^1$, T.~Pierog$^1$, A.~Sabourov$^4$, \\
                      &              &
   T.~Sako$^5$, B.~Stokes$^{2,3}$, R.~Ulrich$^1$ \\
\noalign{\smallskip}\hline\noalign{\smallskip}
 & & 1:~Auger, 2:~HiRes, 3:~TA, 4:~Yakutsk, 5:~LHCf \\
 & & *:~KEK Theory Center \\
\end{tabular}
\end{table}

The WGs addressed such questions as:
how much common understanding have we reached in  UHECR
observations, what are the remaining differences among
experiments, and what are the charges and challenges
to resolve the issues. 
Each WG reported the results of their research in the symposium,
and reports are included in the proceedings
\cite{wg_spectrum} \cite{wg_composition}
\cite{wg_anisotropy} \cite{wg_multimessenger} \cite{wg_simulation}.
Two summary talks were given at the end:
one for the theory and phenomenology by {\it A.~Olinto}, and
another for the experiment and future prospects by {\it M.~Fukushima}
(this report). The symposium  concluded with remarks by {\it A.~A.~Watson} and
a general discussion among all participants.

\section{Energy Spectrum}
\label{sec:spectrum}
The spectrum WG reviewed the energy spectrum of
currently operating experiments, Auger, TA and Yakutsk, and
of completed experiments, AGASA and HiRes.
First, all spectra were fitted with broken power laws. 
The strong suppression of the flux, expected from the GZK process,
was confirmed with good statistical significance
by HiRes, TA and Auger around 10$^{19.6\pm0.2}$ eV, showing
a decrease of the power law index from $-2.7$ down to
$-5.2 \sim -4.2$.
This suppression was not seen in the AGASA spectrum,
and Yakutsk currently has insufficient
exposure in this energy region.
A dip structure, or ``ankle'', was observed by all experiments
as a change of power law index
from $-3.3$ to $-2.7$ around 10$^{18.7}$ eV.

\begin{figure}[t]
  \resizebox{0.90\columnwidth}{!}{
    \includegraphics{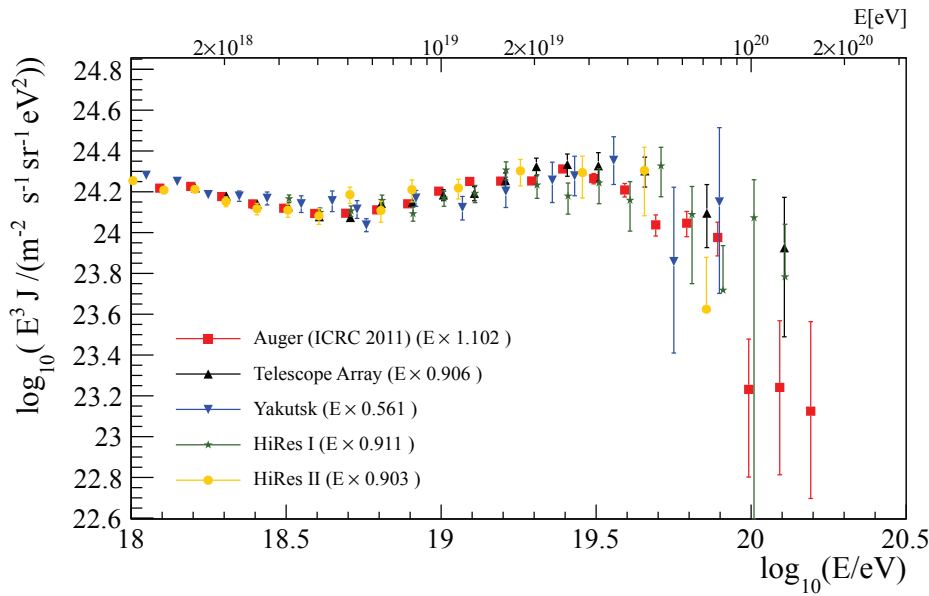} }
    \caption{Re-scaled energy spectra in the form of E$^3$
    $\times$ flux. See spectrum WG's report for details
    \cite{wg_spectrum}.}
    \label{fig:rescaled_spectra_e3}
\end{figure}

Second, the WG examined the consistency of
spectral shapes between experiments
by adjusting the energy scale of each experiment
to a common scale, which was taken halfway between
Auger and TA. This particular choice of common energy scale,
a mean of Auger and TA, reflects our belief that
the energy determination by FD is experimentally better controlled
than SD without  relying too much on the result of EAS simulations
and their hadronic interaction models at UHE. 
Fig. \ref{fig:rescaled_spectra_e3}
was obtained by fitting the spectra below 10$^{19.5}$ eV
to the common energy scale.
Corresponding scale factors
are listed in Table-\ref{tab:energyscale} \cite{wg_spectrum}.
\begin{table}
\begin{center}
  \caption{Energy scale factors obtained by fitting the spectrum \cite{wg_spectrum}.}
  \label{tab:energyscale}
    \begin{tabular}{cccccc}
    \hline\noalign{\smallskip}
    Auger & TA    & AGASA & Yakutsk & HiRes-I & HiRes-II \\    
    1.102 & 0.906 & 0.652 & 0.561   & 0.911   & 0.903    \\
    \noalign{\smallskip}\hline
    \end{tabular}
\end{center}
\end{table}

The estimated energy scale uncertainties of FD measurement
are 17\%, 22\% and 21\% for HiRes, Auger and TA respectively,
and the obtained scale factors are within these limits.
The AGASA scale factor (0.652)
is at the limit of consistency with the quoted
systematic uncertainties of AGASA (20\%) and FD
($\sim$20\% typical).
AGASA's number of events above 10$^{20}$ eV
becomes two after the energy rescaling, and the observation
of the extended spectrum beyond the GZK cutoff becomes
statistically insignificant. The 
Yakutsk SD spectrum is calibrated by the Cherenkov
light emission and its rescale factor (0.561) seems to be 
also at the limit of consistency with the 
systematic uncertainty of the
Cherenkov calibration ($\sim$25\%).

\section{Models}
\label{sec:models}
Observed energy spectra are interpreted by models 
of CR generation and propagation. In those models, the generation
of CRs at the source is usually characterized by a power law
spectrum and the maximum energy of acceleration.
The propagation of CR causes energy losses by collisions 
with the CMB and Extra-galactic Background Light (EBL). In some models,
effects of galactic and extra-galactic magnetic fields are
considered.

\begin{figure}[ht]
  \begin{tabular}{cc}
    \begin{minipage}[b]{0.48\columnwidth}
      \begin{center}
        \resizebox{1.0\columnwidth}{!}{
          \includegraphics{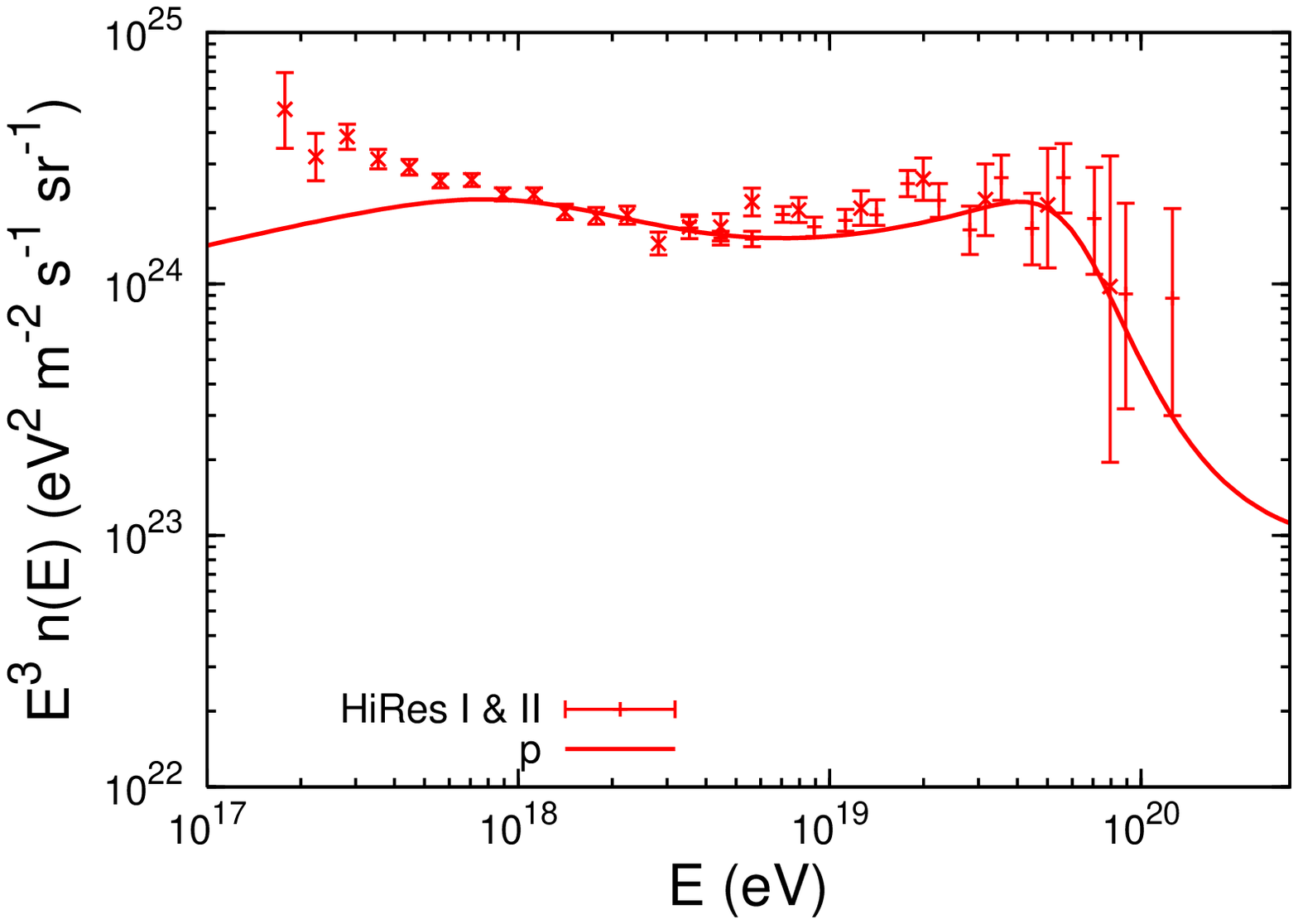} }
        \caption{HiRes spectrum and the prediction of the dip model
        ({\it R.~Aloisio})}
        \label{fig:aloisio_dip}
      \end{center}
    \end{minipage}
    \hspace{3mm}
    \begin{minipage}[b]{0.48\columnwidth} 
      \begin{center}
        \resizebox{1.0\columnwidth}{!}{
          \includegraphics{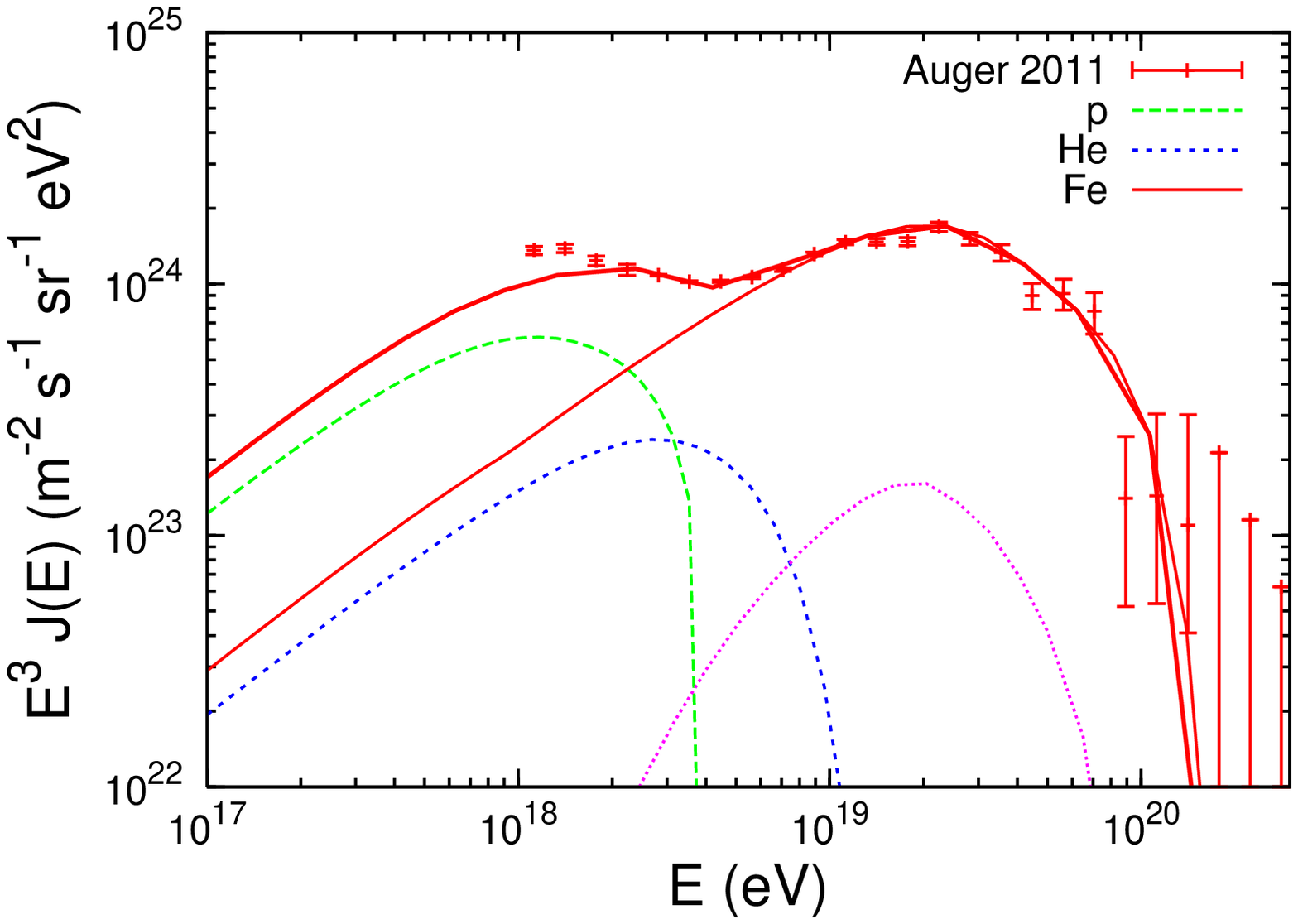} }
        \caption{Auger spectrum and prediction by one of
        the multi-composition models ({\it R.~Aloisio}).}
        \label{fig:aloisio_dis}
      \end{center}
    \end{minipage}
  \end{tabular}
\end{figure}

\noindent
{\bf Dip Model \hspace{3mm}} The observed
energy spectra of HiRes and TA above 
10$^{18}$ eV, in terms of the spectral shapes and
their break point energies, are well explained
by the GZK process;  interactions of extra-galactic protons
with the CMB producing 
the ``ankle'' via e$^+$ e$^-$ pair production at around
10$^{18.7}$ eV, and the ``cutoff'' via pion photo-production
around 10$^{19.6}$ eV (see Fig. \ref{fig:aloisio_dip}
by {\it R.~Aloisio}).  
The energy E$_{1/2}$ of the integrated flux
is 10$^{19.73\pm0.07}$ eV for the HiRes spectrum, and 
agrees with the theoretical calculation
of 10$^{19.72}$ eV ({\it V.~Berezinsky}) within a systematic uncertainty
of the energy scale of HiRes (17\%).
This picture (dip model) is consistent with the observation of a
proton-like composition above 10$^{18}$ eV by HiRes and TA using
the X$_{max}$ measurement.

\smallskip\noindent
{\bf Multi-Composition Model \hspace{3mm}} 
On the other hand, Auger, with its measured composition
changing from proton to heavier nuclei in this energy range,
has not associated the observed spectral features
with a particular physics mechanism. A variety of models
with different source composition, acceleration and propagation
assumptions have been proposed to fit Auger's high statistics
spectrum data, and measurements of $<X_{max}>$ and RMS(X$_{max}$).
See Fig. \ref{fig:aloisio_dis} for example for the case
of p-He-Fe multi-composition by {\it R.~Aloisio}. 
Although these models often contain CR acceleration limits 
at the origin as important ingredients, it is generally
conceived that the photo-disintegration of iron nuclei
by the CMB is the dominant cause creating the high energy
flux suppression. The effect of photo-disintegration on nuclei
was already mentioned in the original GZK papers. 
The ankle may be explained in these models as the acceleration
limit of extra-galactic protons, or of the galactic heavy components
corresponding to a galactic to extra-galactic transition of CRs.

The following experimental challenges are identified to pin down the
correct physics model of UHECR spectrum:
\begin{itemize}
\item
{\bf Energy scale uncertainty in FD: }
The flux suppression by the GZK process, whether it is for proton or
for iron, is absolute in energy scale, and the parameter such as
E$_{1/2}$ is relatively insensitive to the choice of the spectral
index at the source and the cosmological evolution of CRs.
Note that, between Auger and TA, there is a significant difference
in the energy scale; Auger is $\sim$18\% lower than
TA in the rescaling of energy spectra, and Auger's cutoff breakpoint energy
is $\sim$46\% lower than that of TA (10$^{19.68}$ eV for TA and 10$^{19.41}$ eV
for Auger. See \cite{wg_spectrum}). 

\item
{\bf Energy linearity in SD: }
The calibration of SD energy is made only by the limited sample
of SD/FD coincident events, i.e. the direct calibration is limited
to energies below the onset of the high-energy cutoff. The extrapolation
to higher energies relies on the validity of the
Constant Intensity Cut (CIC) method for Auger, and Monte Carlo air
shower simulation for TA, both of which require independent
experimental verification. 
\item
{\bf Mass Composition: } The mass composition plays a key role
in distinguishing different models. The change of $<X_{max}>$ also
directly affects the energy determination. 
Meaningful measurement of X$_{max}$ is now limited to 10$^{18.0}$ eV
$-$ 10$^{19.5}$ eV. An extension of composition
measurement to both higher and lower energy regions is important.

\item
{\bf Around the cutoff and above: }
The spectral shape above 10$^{19.5}$ eV may be
a convolution of the GZK cutoff
and the acceleration limit of UHECRs. The cutoff shape
also carries information on the composition as seen
in Figs.\ref{fig:aloisio_dip} and \ref{fig:aloisio_dis}.
Above 10$^{19.5}$ eV, the agreement of spectral shapes among experiments
is not as satisfactory as the lower energy region (see
Fig.\ref{fig:rescaled_spectra_e3}).
In order to determine the exact shape
of the cutoff above 10$^{19.5}$ eV,
measurements with higher statistics and less systematic error
are essential.
 
\end{itemize}

\section{Mass Composition}
\label{sec:composition}
The primary mass composition of UHECRs is determined from the
X$_{max}$ measurement. The X$_{max}$ can be determined for each FD event,
but the mass composition is usually determined only statistically
by comparing the $<X_{max}>$ and RMS(X$_{max}$) with the expectation from
the EAS simulation. This is due to the large event-by-event fluctuation
of X$_{max}$, particularly for proton events, and the limited accuracy of X$_{max}$
measurement. The most straight-forward way of measuring X$_{max}$ is by 
imaging the EAS development in the atmosphere using 
the FD, which has been adopted by HiRes, Auger and TA. 
There are less direct way of measuring the Cherenkov light distribution
on the ground (by Yakutsk) and muon detection on the ground (by Auger/SD). 

\begin{figure}[ht]
  \resizebox{1.02\columnwidth}{!}{
  \includegraphics{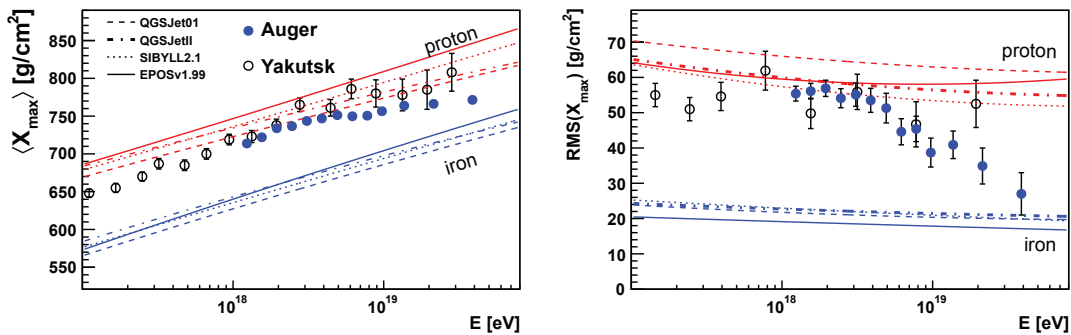} }
  \caption{The energy dependence of $<X_{max}>$ for Auger and Yakutsk (left).
  The measured $<X_{max}>$ for Auger and Yakutsk has
  negligible analysis biases, and is directly comparable with
  the simulation (red and blue lines). 
  The energy dependence of RMS(X$_{max}$) for Auger and Yakutsk is shown on the right.
  The RMS(X$_{max}$) of Auger is defined as the r.m.s. of the X$_{max}$ deviations.
  The experimental resolution estimated from the simulation
  is subtracted in quadrature. The highest energy bin of Yakutsk
  contains only 3 events.}
  \label{fig:auger_yakutsk_xmax}
\end{figure}

\begin{figure}[h]
  \begin{tabular}{cc}
    \begin{minipage}[b]{0.48\columnwidth}
      \begin{center}
        \resizebox{0.96\columnwidth}{!}{
          \includegraphics{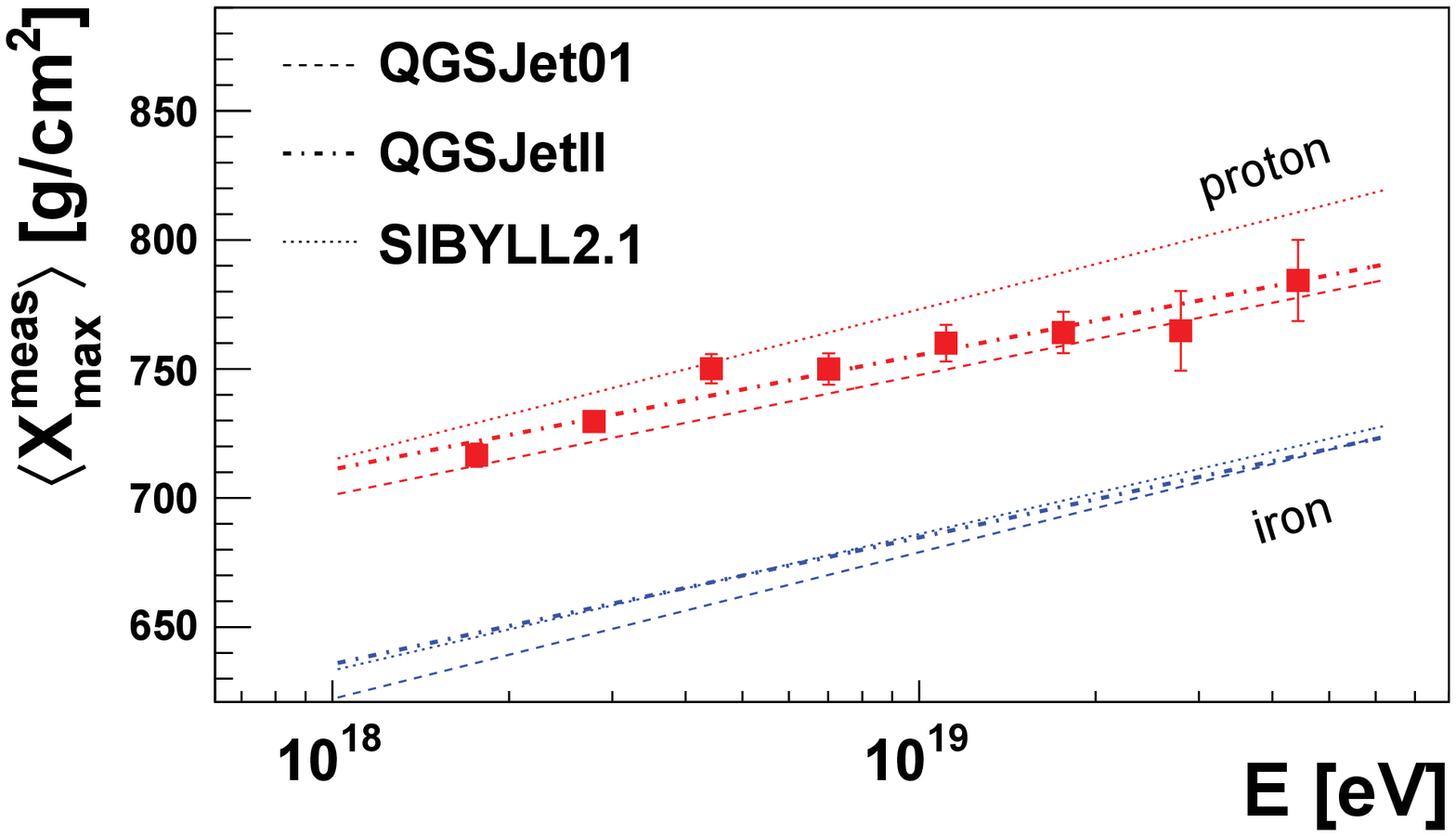} }
          \caption{The energy dependence of $<X_{max}>$ for HiRes.
          The measured $<X_{max}>$ is compared with the simulation
          (lines) with analysis biases included both for the measurement
          and simulation.}       
        \label{fig:hires_xmax}
      \end{center}
    \end{minipage}
    \hspace{3mm}
    \begin{minipage}[b]{0.48\columnwidth} 
      \begin{center}
        \resizebox{0.96\columnwidth}{!}{
          \includegraphics{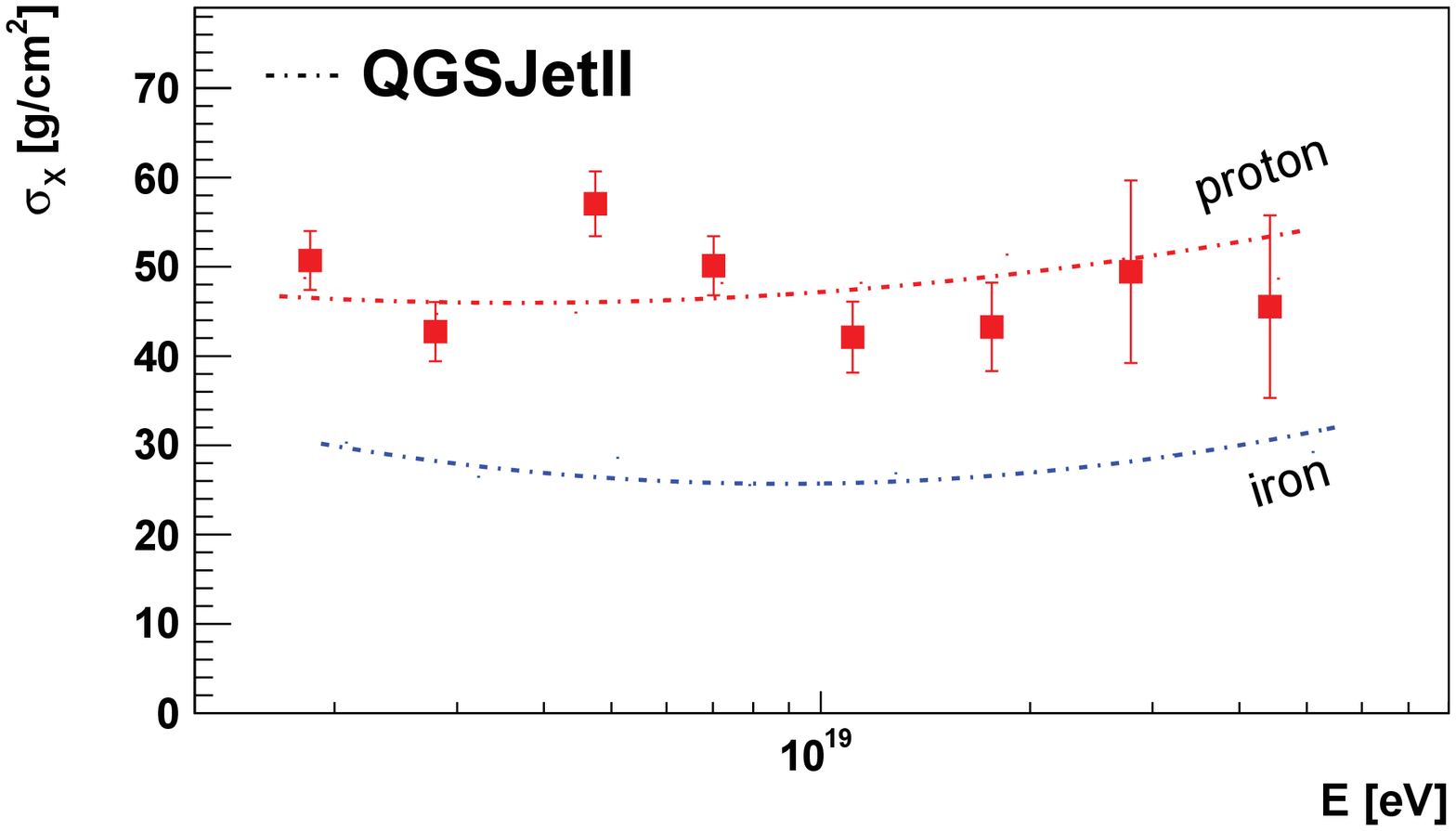} }
        \caption{The energy dependence of RMS(X$_{max}$) for HiRes.
        The RMS(X$_{max}$) is defined as the $\sigma$ obtained by fitting
        the X$_{max}$ distribution with a Gaussian truncated at 2$\sigma$. 
        The experimental resolution is included for both 
        the measurement and simulation.}
        \label{fig:hires_xmax_sigma}
      \end{center}
    \end{minipage}
  \end{tabular}
\end{figure}

\begin{figure}[h]
  \begin{tabular}{cc}
    \begin{minipage}[b]{0.45\columnwidth}
      \begin{center}
        \resizebox{0.98\columnwidth}{!}{
          \includegraphics{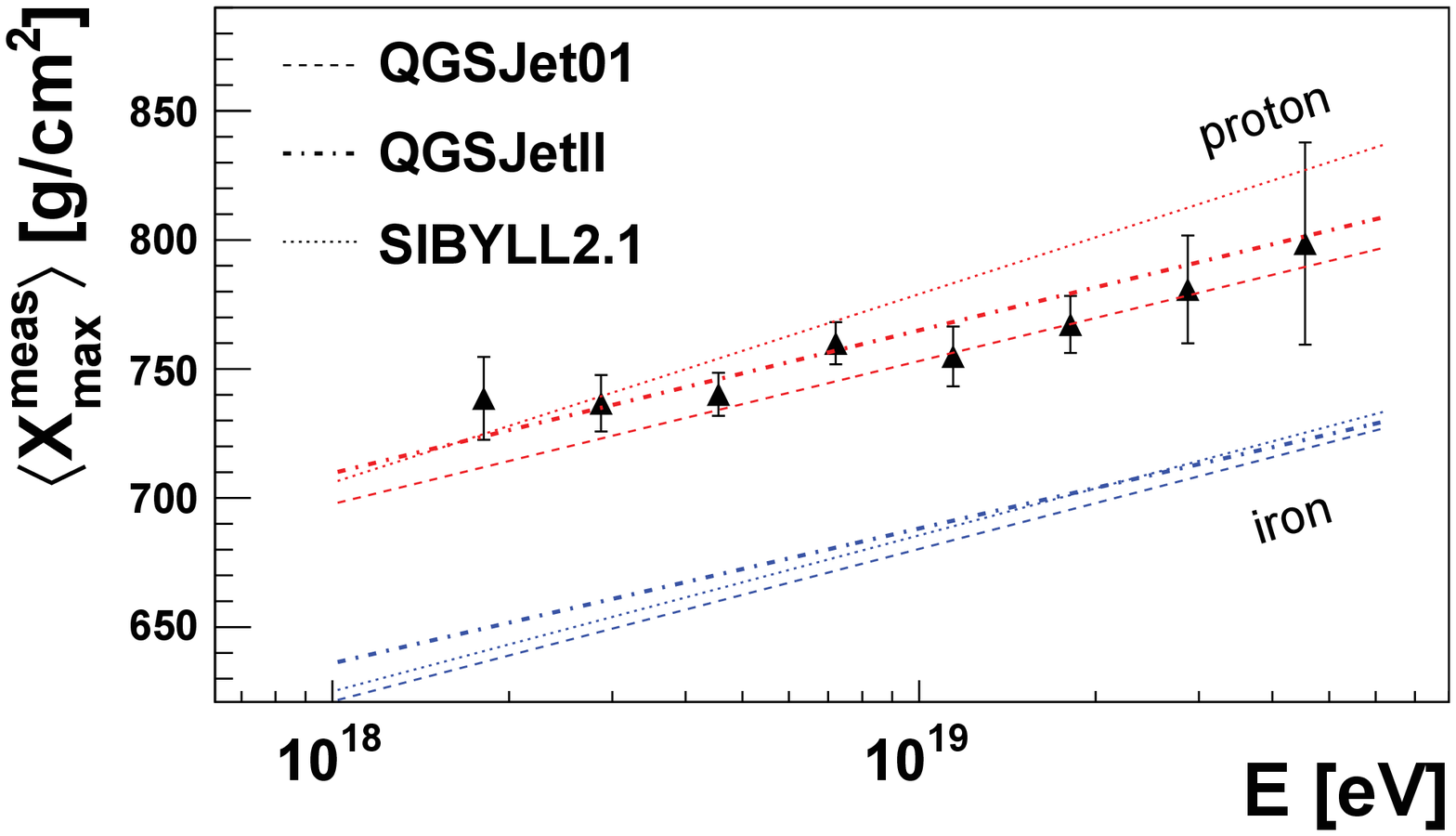} }
        \caption{The energy dependence of $<X_{max}>$ for TA (preliminary).
         The measured $<X_{max}>$ is compared with the simulation
         (lines) with analysis biases included for both the measurement
         and simulation.}
        \label{fig:ta_xmax}
      \end{center}
    \end{minipage}
    \hspace{3mm}
    \begin{minipage}[b]{0.50\columnwidth} 
      \begin{center}
        \resizebox{1.0\columnwidth}{!}{
          \includegraphics{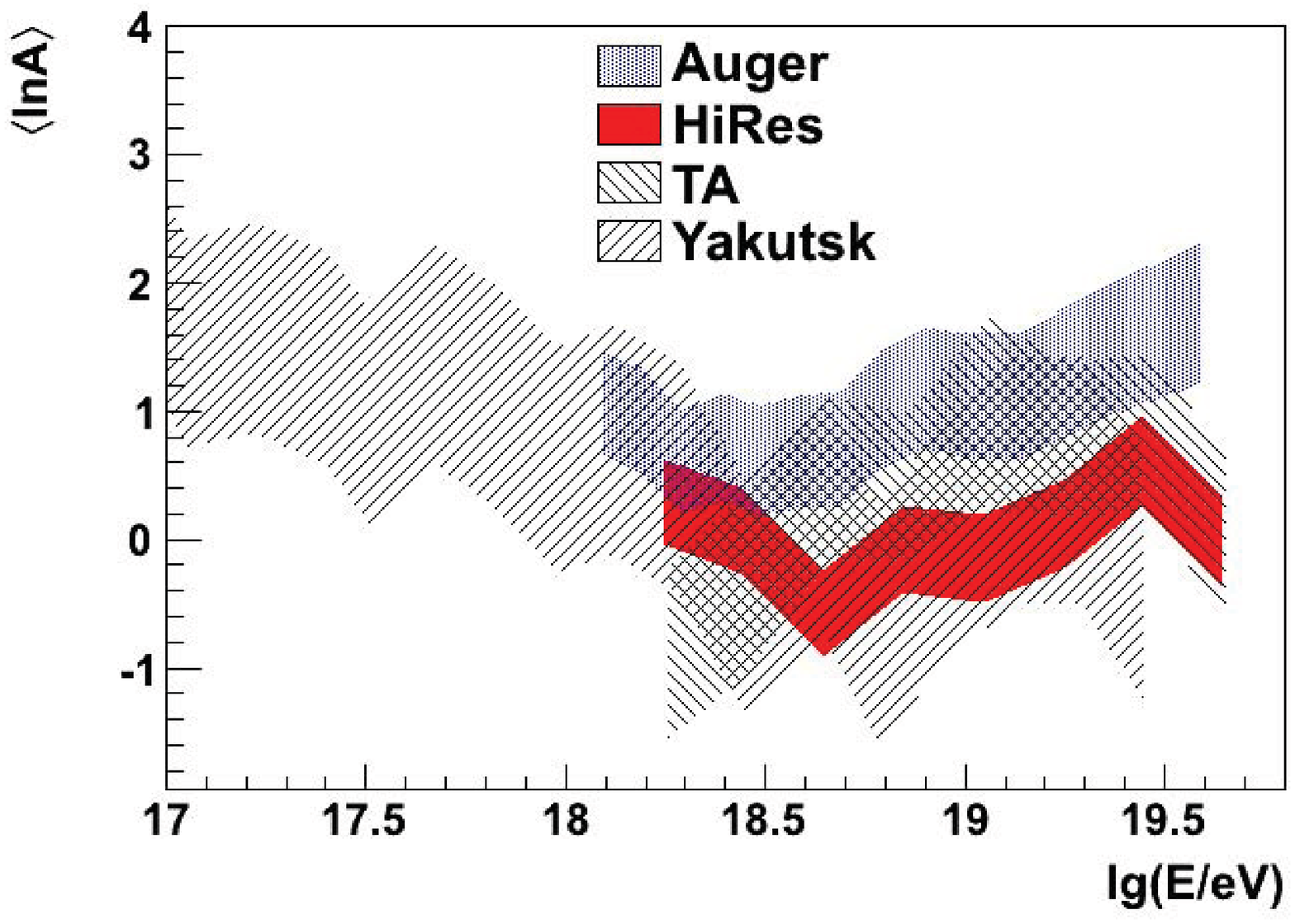} }
          \caption{The energy dependence of average composition $<$lnA$>$ using
          QGSJET-II. The shaded area indicates
          systematic uncertainties of each experiment.
          The energies are rescaled as in the case for the spectrum comparison.}
        \label{fig:lna_comparison}
      \end{center}
    \end{minipage}
  \end{tabular}
\end{figure}

Measurements of $<X_{max}>$ and RMS(X$_{max}$) 
are compiled by the composition WG \cite{wg_composition}. 
They are shown in Fig. \ref{fig:auger_yakutsk_xmax} for Auger and Yakutsk,
in Figs. \ref{fig:hires_xmax} and \ref{fig:hires_xmax_sigma} for HiRes, 
in Fig. \ref{fig:ta_xmax} for TA (preliminary)
together with the results of air shower simulations for protons and iron.
The different measurements are not overlaid in one plot because 
each experiment treats, either includes or removes, analysis biases
and resolution effects differently, and no direct comparison of
measured values is adequate.  

As seen in the plots,
all experiments agree that the light component (proton) dominates
in the energy region below the ankle
for 10$^{18.2}$ eV $-$ 10$^{18.5}$ eV.
Above 10$^{18.5}$ eV, Auger's measurement suggests
a change of composition from light to heavier,
reaching almost iron at 10$^{19.5}$ eV. This is seen
in both $<X_{max}>$ and RMS(X$_{max}$), but is
more pronounced in RMS(X$_{max}$).
On the other hand, no clear indication of the composition change
is seen for the $<X_{max}>$ of HiRes, TA and Yakutsk
up approximately to 10$^{19.3}$ eV.

Distinctions between Auger and HiRes/TA/Auger is obviously
that Auger covers the southern sky,
and HiRes/TA/Yakutsk cover the northern sky.
Among other important distinctions are event detection topologies
and data analysis methods;
Auger uses hybrid events using FD and SD (muon based) in coincidence,
and HiRes and TA use stereo FD events using two FDs (EM energy based)
in coincidence. In the data analysis, Auger uses tight acceptance cuts 
to minimize the analysis bias, whereas HiRes and TA use looser acceptance
cuts to accept more events and use elaborate simulations to
estimate the acceptance bias. The composition WG scrutinized
the analysis methods of Auger, HiRes and TA
in order to understand whether the analysis method 
is responsible for the difference.

In order to guarantee good measurements 
of X$_{max}$, all experiments apply the so-called
bracket cut for the acceptance. This cut requires that the X$_{max}$ of
individual events be observed in the Field of View (FoV) of the
FD telescope. The tight bracket condition, however, causes
certain bias in $<X_{max}>$ and RMS(X$_{max}$) because,
for a given event geometry, we do not know {\it a priori} how many
unbracketed events exist outside the FoV. 

HiRes and TA developed an elaborate EAS event simulation, 
analyzed the generated MC events using 
the same calibration and analysis programs as the data,
and the data-MC agreement was checked. The event reconstruction was
made by the inverse-MC method using the same parameter set
for the data and MC.
The data observables, including $<X_{max}>$ and RMS(X$_{max}$)
with its energy dependence, are well reproduced 
by the pure proton primary using the QGSJET-II interaction model,
and this is the basis for the claim of an unchanged proton component
by HiRes and TA.
The estimated acceptance bias of TA is less than
15 g/cm$^2$ for QGSJET-II proton
above 10$^{18.6}$ eV, and is negligible for QGSJET-II iron. 
The reconstruction bias, defined as the difference between the real
(thrown) X$_{max}$ and the reconstructed X$_{max}$, is approximately 10 g/cm$^2$
independent of the energy and the composition. 

Auger's X$_{max}$ analysis requires tighter condition for the acceptance
such that an event is accepted only if not only the observed X$_{max}$
but also an entire range of the probable X$_{max}$
of that event geometry (and energy) is included in the FoV.
The range of accepted X$_{max}$ is determined by checking that the resultant
$<X_{max}>$ is not influenced by the FoV cuts \cite{wg_composition}.
This method is tuned to minimize the X$_{max}$ acceptance bias, at the cost of
fewer accepted events; the tight FoV cut reduces the accepted events
of Auger by half for energies above 10$^{18.2}$ eV. The performance
of the analysis was estimated by using the MC; the combined
(reconstruction + acceptance) X$_{max}$ bias is smaller than
4 g/cm$^2$, and the X$_{max}$ resolution is 20$-$25 g/cm$^2$.

The composition WG made numerous checks on the analysis results, and
the data-MC agreement was examined for many observables as
reported in \cite{wg_composition}.
In short, no particular flaw of analysis, or definite clue
for the resolution of the difference
was identified so far. The WG is now proceeding 
to the next step to analyze the Auger data with TA's looser cuts, 
to analyze the TA data with Auger's tight cut, in order 
to see whether the result is unchanged. Work is also in progress 
to generate and analyze MC event sets 
having the same X$_{max}$ distribution as observed by Auger, and
pass them through the analysis chain of TA and HiRes. 

The results of $<X_{max}>$ from different experiments
(or analysis methods) may be compared
by converting the $<X_{max}>$ into $<$lnA$>$ \cite{wg_composition},
as it is known that the change of X$_{max}$ is approximately
proportional to lnA at a fixed CR
energy, where A is the atomic number of the primary CR.
The composition WG calculated the value of $<$lnA$>$ for each experiment
by proportionating the difference of the measured X$_{max}$ between
the expected X$_{max}$ for protons and iron. 
The result obtained by using QGSJET-II is shown in Fig. \ref{fig:lna_comparison}.
The shaded area indicates the systematic errors of each experiment;
12 g/cm$^2$ for Auger and TA, 20 g/cm$^2$ for Yakutsk, 6 g/cm$^2$ for HiRes.
It is seen that HiRes and Auger give different results and are 
not compatible for energies above 10$^{18.5}$ eV.
The results of TA and Yakutsk partially overlap both with HiRes and Auger.
A comparison using the SIBYLL model makes the difference between Auger and HiRes
smaller, but they do not overlap above 10$^{18.5}$ eV.

It should be noted that the statistics of Auger is
higher than all other experiments combined, even with the
tight acceptance cut
\footnote{The number of events above 10$^{19}$ eV is 452 for Auger,
123 for HiRes, 67 for TA and 22 for Yakutsk.}.
The WG report demonstrated that the energy dependence of Auger's $<$lnA$>$
is statistically not consistent with a constant value of $<$lnA$>$,
whereas the same plots by other experiments are statistically
consistent with both constant and changing values of $<$lnA$>$
with respect to log(E) (see Fig.12 of \cite{wg_composition}).
More events (preferably $\times$3 or more) in future,
particularly in the northern hemisphere, are 
needed to conclude definitively whether the composition
stays light as the energy increases, or turns heavy
as in the case of Auger in the southern hemisphere.   
In addition, on-going improvements of EAS simulation
with better models and newly available collider data
(see section \ref{sec:simulation}) are expected
to contribute a lot to predict the $<X_{max}>$, RMS(X$_{max}$)
and its energy development in a more reliable way.
The composition WG will continue its activity in future.

Finally, the FD and SD hardware and their calibration
needs due attention for the X$_{max}$ comparison. This
becomes important when the comparison of
analyses does not solve the problem.
HiRes, TA and Auger use different optics, PMT matrix
and signal recording methods for the FD telescope. 
The atmospheric models and its attenuation effects are
different for each experimental site.   
{\it S.~Ogio,} in his talk of future plans \cite{future_emcal},
proposed to exchange a set of SDs and FDs between Auger and TA sites,
and to make a simultaneous observation of the same UHECR event.
This will be very effective in excavating systematics in the
detector and analysis, if any important one exists.

\section{Anisotropy}
\label{sec:anisotropy}
\begin{figure}[ht]
  \begin{tabular}{cc}
    \begin{minipage}[b]{0.48\columnwidth}
      \begin{center}
        \resizebox{1.0\columnwidth}{!}{
          \includegraphics{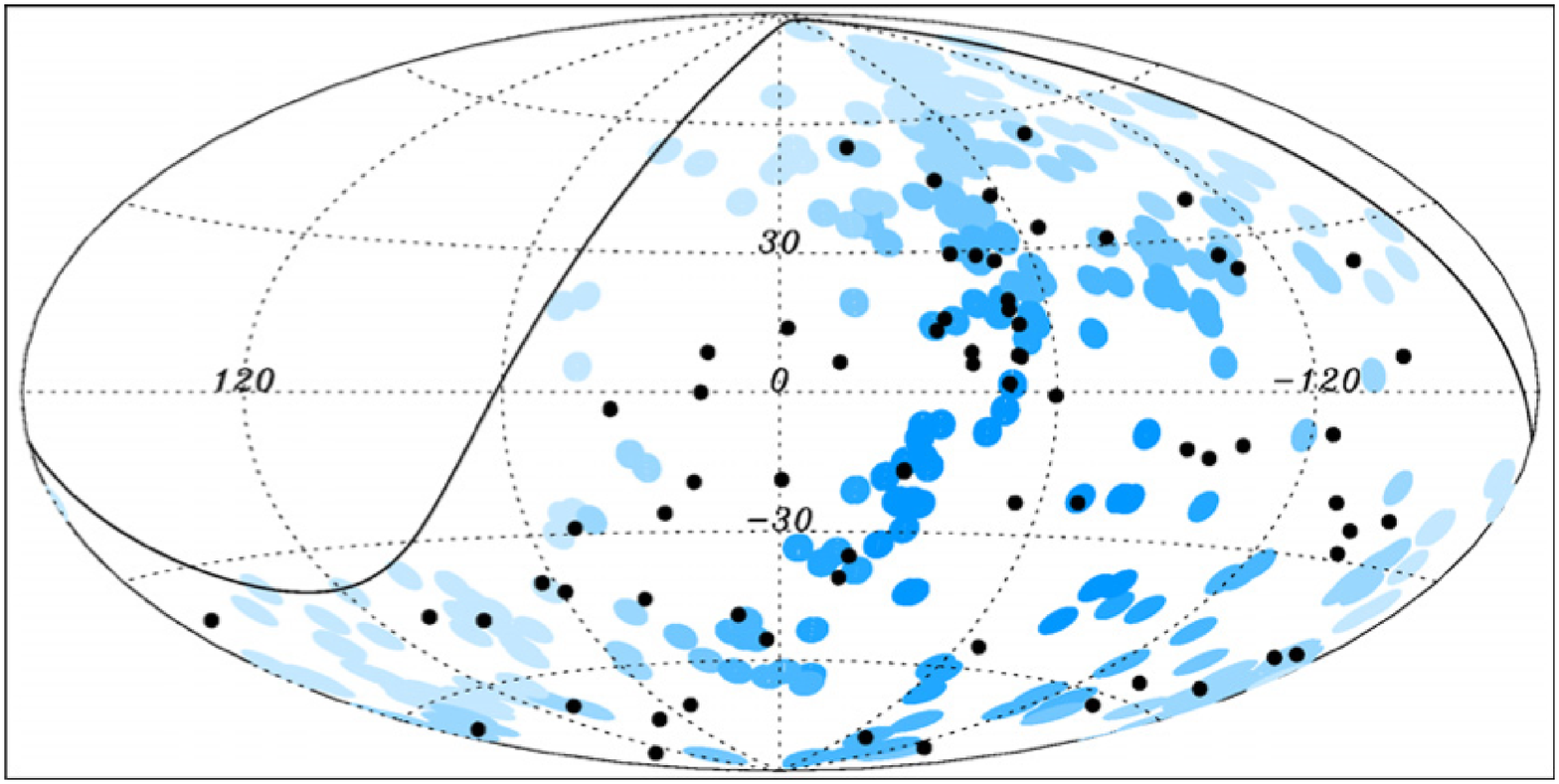} }
        \caption{Arrival directions of UHECRs (E $>$ 57 EeV) in the southern sky
         observed by Auger. A correlation with nearby AGNs
         (z $<$ 0.018) was suggested.}
        \label{fig:vcv_sky_auger}
      \end{center}
    \end{minipage}
    \hspace{3mm}
    \begin{minipage}[b]{0.48\columnwidth} 
      \begin{center}
        \resizebox{0.95\columnwidth}{!}{
          \includegraphics{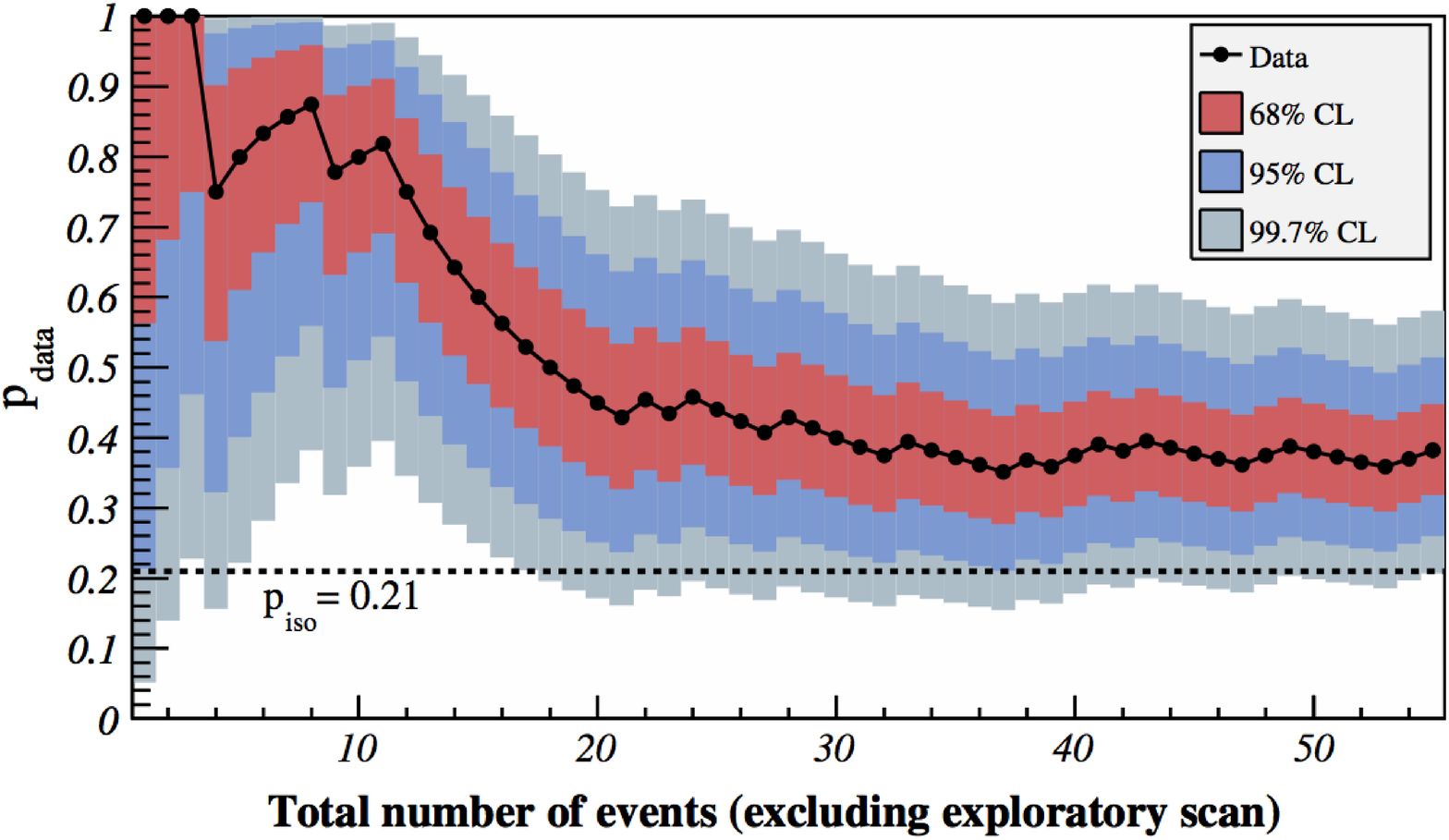} }
        \caption{Time development of Auger AGN correlation. The ratio of
        AGN-correlated events over the total is plotted. The dashed line
        corresponds to the ratio expected from the accidental coincidence.}
        \label{fig:vcv_time_auger}
      \end{center}
    \end{minipage}
  \end{tabular}
\end{figure}

It is expected that the magnetic deflections of UHECRs
by the galactic and extra-galactic magnetic fields
decrease down to a few degrees for energies larger
than 10$^{20}$ eV, and a correlation between the UHECR arrival
direction and the source ``star'' would be showing up. 
This correlation would be particularly  distinct if the
source astrophysical objects are limited within the GZK horizon
($\sim$100 Mpc). In 2007, Auger published a correlation of UHECR 
(E $>$ 57 EeV
\footnote{57 EeV = 10$^{19.76}$ eV, EeV = 10$^{18}$ eV. The original
publication had a threshold energy of 56 EeV. It was later stated
to be equivalent to 57 EeV in the updated energy scale.
}) arrival directions
and nearby AGNs (z $<$ 0.018, or within 75 Mpc) in the
Veron-Cetty and Veron (VCV) catalog, 12th edition
within a separation
of 3.1$^\circ$ \cite{auger_vcv}. 
The result was updated in the Beijing ICRC conference
and its skymap and time development diagram 
are shown in Figs. \ref{fig:vcv_sky_auger}
and \ref{fig:vcv_time_auger}. The rate
of correlated UHECR events presented in Beijing
decreased from 61\% to 33\% but it is still $\sim$3.0$\sigma$ (P=0.006)
away from a chance correlation.
A concentration of correlated events in the vicinity
of Cen A was reported in \cite{auger_cenA},
but the significance did
not increase despite higher statistics
\cite{auger_kh_beijing}.

\begin{figure}[h]
  \begin{tabular}{cc}
    \begin{minipage}[b]{0.48\columnwidth}
      \begin{center}
        \resizebox{0.96\columnwidth}{!}{
          \includegraphics{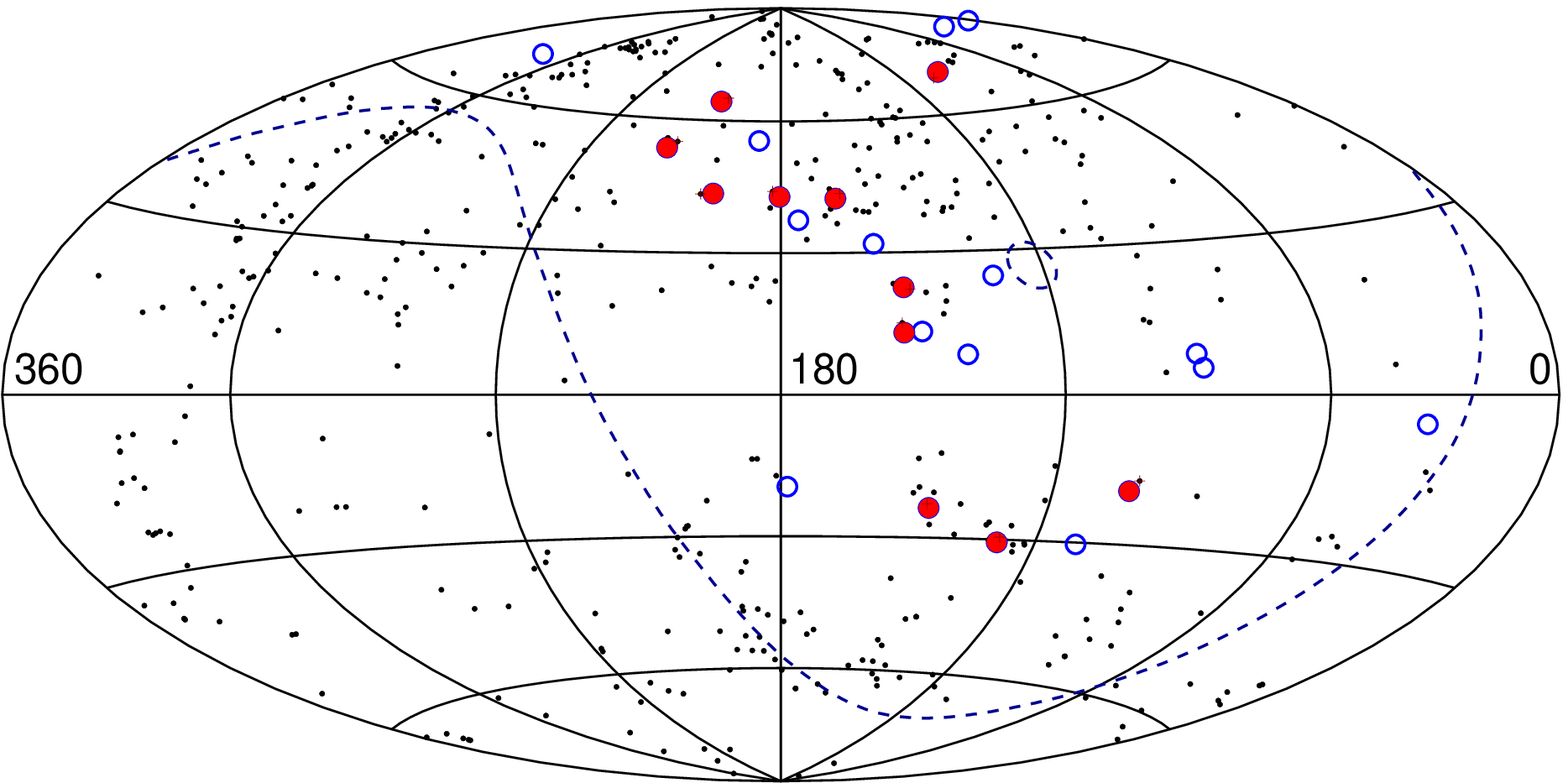} }
        \caption{Arrival directions of UHECRs in the northern sky
         observed by TA. The same correlation conditions
         as in Fig. \ref{fig:vcv_sky_auger} are applied.}
        \label{fig:vcv_sky_ta}
      \end{center}
    \end{minipage}
    \hspace{3mm}
    \begin{minipage}[b]{0.48\columnwidth} 
      \begin{center}
        \resizebox{0.90\columnwidth}{!}{
          \includegraphics{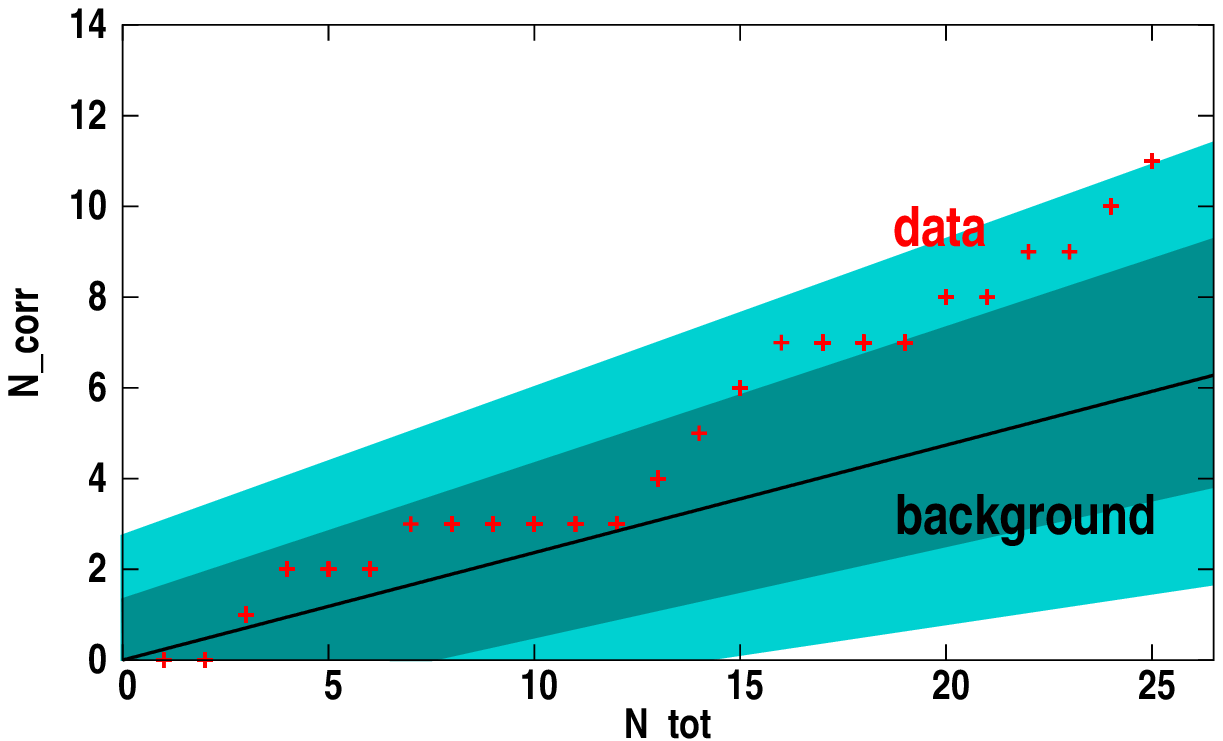} }
        \caption{Time development of AGN correlation of TA. The number of 
        AGN-correlated events is plotted. The solid line
        corresponds to the number expected from the accidental coincidence.}
        \label{fig:vcv_time_ta}
      \end{center}
    \end{minipage}
  \end{tabular}
\end{figure}

HiRes searched for a correlation in the northern
hemisphere but did not find any.
TA searched for correlations in the recent SD data sample with zenith
angles less than 45$^\circ$, and found
11 events are correlated from the total of 25 events
under the same conditions as the Auger AGN correlation \cite{ta_anisotropy}
\footnote{The energy scales of TA/HiRes may be $\sim$20\%
larger than that of Auger as seen in section \ref{sec:spectrum},
and the effective energy threshold for the
search may be different by this amount.}.
The number of correlation expected from the uniform and random set
of 25 events is 5.9, or the observed ``correlation'' is more than
2$\sigma$ (P=0.02) away from the accidental correlation. Both HiRes
and TA did not optimize 
the search parameters and used the same conditions
as Auger in order to stay free from the search bias.
The skymap of TA and the time development of the correlation are
shown in Figs. \ref{fig:vcv_sky_ta} and \ref{fig:vcv_time_ta}.
 
Auto-correlations among UHECR events are searched for by HiRes, Auger and TA
but no positive results have been reported.
A global harmonic analysis was performed by Auger, and correlations with the 
Large Scale Structure (LSS) of galaxies have been searched for by HiRes, Auger and TA,
but the results are so far consistent with both the isotropy and the LSS. 

The anisotropy WG summarized that, despite a large effort spent
on the search for anisotropy and correlation,
no clear signal has so far been established
with certainty \cite{wg_anisotropy}.
Various hints, however, are emerging from the observations of Auger and TA,
and they should be closely followed in coming days. 
The prospects and challenges by the anisotropy WG are
reiterated as follows:
\begin{itemize}
\item
{\bf Composition: }
Expectation for correlation obviously depends on the actual
composition, or the atomic number, of the primary UHECRs. 
It should be noted that above 10$^{19.7}$ eV where
the correlations with AGNs and LSS are expected,
no direct measurement of composition by X$_{max}$ exists.
Discrimination power for composition on an event-by-event
basis would be a great asset for the future anisotropy search. 

\item
{\bf Large Exposure: }
The next 5 years will be an exciting time when some of the
hints of anisotropy seen by Auger and TA
will be clarified with more statistics. 
In order to fully test the observed anisotropy,
or to find new kind of anisotropy and characterize its cause,
one order of magnitude larger exposure is needed by the next generation
observatory.

\item
{\bf Energy Assignment: }
An accurate energy assignment of each event is essential for making energy cuts
for trans-GZK anisotropy and also for large-scale anisotropy studies
at EeV energies. The energy scale error (of FD)
is expected to be improved from the present $\sim$20\% down to $\sim$10\%,
matching the energy resolution (of SD) at the highest energy region.
The present angular resolution of $\sim$1$^\circ$ may be sufficient
for the future anisotropy study
\footnote{For TA/SD events, the energy resolution is 13\% and the angular
resolution is 1.1$^\circ$ for E $>$ 57 EeV by simulation studies. Other
experiments have similar numbers.}.

\item
{\bf Full-sky coverage: }
There are many advantages to operate an observatory with full-sky coverage;
all possible point sources would be exposed, and the structure of anisotropy
with minimal exposure distortion could be achieved.
Besides, with full-sky coverage, the methods of spherical harmonics
and multipole analysis could be applied without additional assumptions
on the anisotropy behavior in the uncovered region of the sky.
Unified analysis of southern and northern skies
are common interests of existing groups; such analysis was initiated using
the Auger and TA data by the anisotropy WG, and is being pursued now. 

\item
{\bf Wide Energy Coverage: }
Currently preferred models of UHECR (see section \ref{sec:models})
suggest that cosmic rays are extra-galactic above 10$^{18}$ eV,
but the scenario of transition from galactic to extra-galactic
below 10$^{18}$ eV is not established, and the predictions of
models differ (see {\it Berezinsky}).
Understanding this transition is an important step forward
to provide a complete picture for origins of UHECRs.
The transition should be signaled by the composition change (from heavy
to light), spectrum deformation (second knee?) and the energy-dependent
anisotropy as hinted by the phase change in the harmonic
analysis \cite{wg_anisotropy} of Auger.
Such low-energy extensions are proceeding well as HEAT and AMIGA in Auger
and TALE in TA to supply high quality, high statistics data
down to 10$^{16.5}$ eV in the near future.
\end{itemize}

\section{Multi-Messenger (UHE gammas and neutrinos)}
\label{sec:multimessenger}
Any top-down model of UHECR origin expects abundant UHE $\gamma$
rays and neutrinos
as a decay product of $\pi$ mesons. The GZK process also produces
UHE gammas and neutrinos from
pion decays of the excited nucleon resonance. 
No such UHE cosmic $\gamma$'s and $\nu$'s have been observed so far by
the UHECR experiments. 

\begin{figure}[ht]
  \begin{tabular}{cc}
    \begin{minipage}[b]{0.48\columnwidth}
      \begin{center}
        \resizebox{1.03\columnwidth}{!}{
          \includegraphics{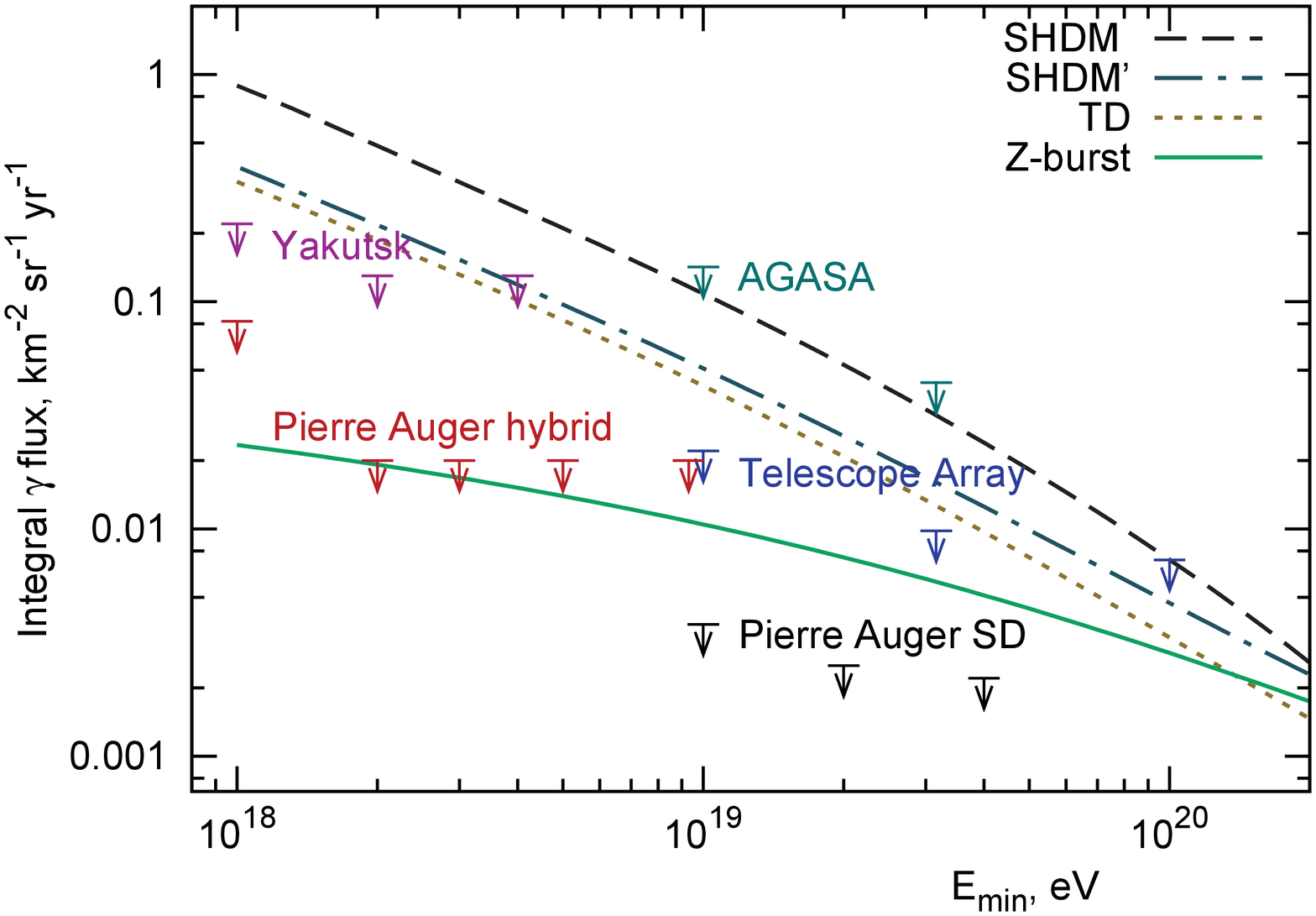} }
          \caption{Experimental limits placed on the UHE photon flux
          and expectation from several top-down acceleration models
          of UHECRs.}
        \label{fig:limit_topdown}
      \end{center}
    \end{minipage}
    \hspace{3mm}
    \begin{minipage}[b]{0.48\columnwidth} 
      \begin{center}
        \resizebox{1.00\columnwidth}{!}{
          \includegraphics{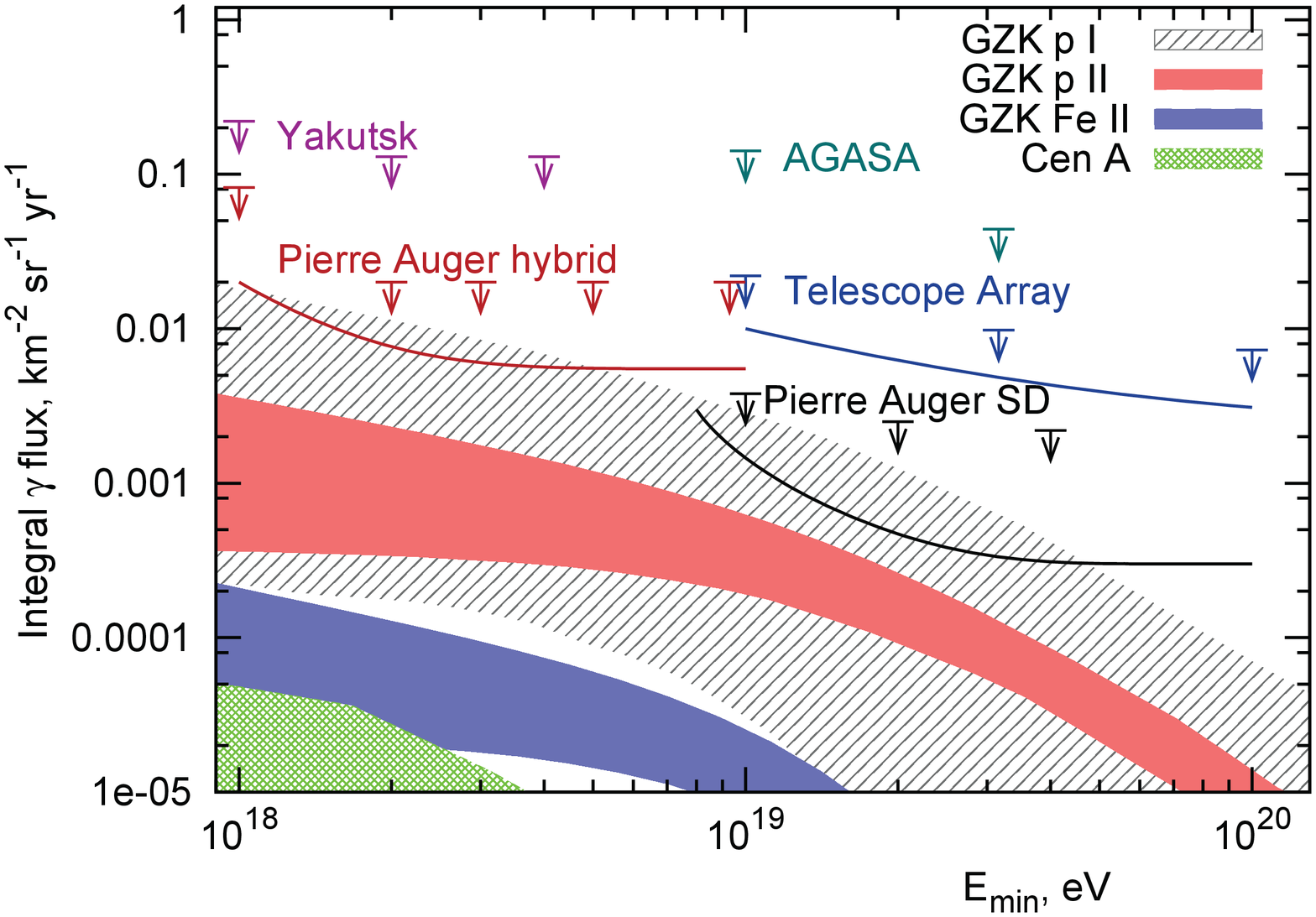} }
          \caption{Experimental limits placed on the UHE photon flux. 
          Expected limits by 2015 are indicated by lines. The shaded region
          corresponds to the flux predicted by several
          cosmogenic models \cite{wg_multimessenger}.}
        \label{fig:limit_gamma}
      \end{center}
    \end{minipage}
  \end{tabular}
\end{figure}

\begin{figure}[t]
  \begin{center}
    \resizebox{0.65\columnwidth}{!}{
      \includegraphics{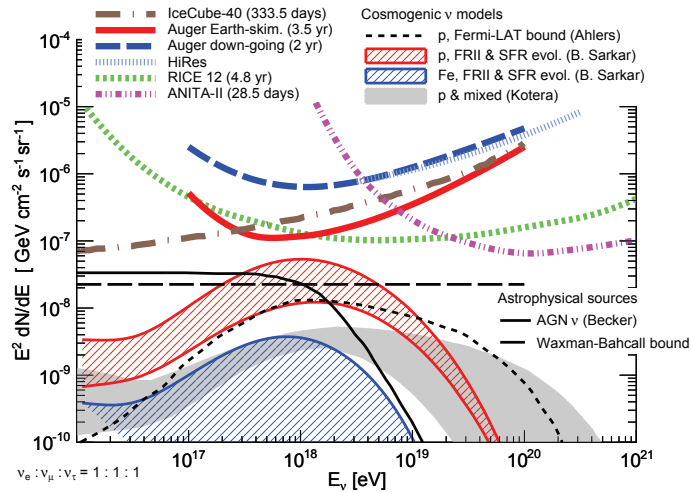} }
      \caption{Experimental limits placed on the UHE neutrino flux.
      Expected fluxes from several cosmogenic models are shown by
      the shaded region and black lines. For details, see
      WG's report \cite{wg_multimessenger}.}
      \label{fig:limit_neutrino}
  \end{center}
\end{figure}

Limits on the flux of UHE $\gamma$'s
by Auger, TA and Yakutsk already place strong constraints on the top-down
models as seen in Fig. \ref{fig:limit_topdown} \cite{wg_multimessenger}.
The multi-messenger WG further compiled the present
(and ``anticipated'' by 2015) experimental upper limits
and compared them with the expected flux from
cosmogenic (GZK) production models.
They are reproduced here in Figs.\ref{fig:limit_gamma}
and \ref{fig:limit_neutrino}.

The expected sensitivities for UHE $\gamma$'s in Fig. \ref{fig:limit_gamma}
tells us that the next generation UHECR detector
with one order of magnitude larger exposure than the present
Auger and TA has a significant chance to detect cosmogenic $\gamma$ rays,
provided the primary composition is proton. The iron primary is expected to
give an order of magnitude smaller flux than the proton primary.
The UHECR detectors planned for future,
such as the nGO and JEM/EUSO (section \ref{sec:futures}) may be the only
detectors to observe UHE cosmogenic $\gamma$ rays.

For the case of cosmogenic $\nu$'s, as seen
in Fig \ref{fig:limit_neutrino}, the 
recently completed IceCube (IC-86) will have a good chance of detecting
UHE $\nu$'s after $\sim$5 years of running in the region of
10$^{17}$ eV $-$ 10$^{19}$ eV, for the case of primary protons. 
The south pole radio experiments, ANITA by balloon and 
RICE in the ice, already contribute to set upper limits for energies
greater than 10$^{19}$ eV. Future huge ground array nGO
with 10 times larger acceptance than the present Auger,
covers the UHE $\nu$ detection in the region of 10$^{17}$ eV $-$ 10$^{20}$ eV,
and in case if detected, it would give most precise information on the
energy and arrival direction, and possibly the species (e, $\mu$, $\tau$)
of the UHE neutrinos. The JEM/EUSO will have the largest sensitivity but
its energy acceptance is limited above 10$^{19.5}$ eV. The iron primary
or the mixed composition may reduce the detection possibility of cosmogenic
neutrinos by roughly an order of magnitude.

\section{Hadronic Interaction and AS simulation}
\label{sec:simulation}
Despite the success of determining the UHECR energy by the
new air fluorescence technique, and identifying the GZK traces
in the energy spectrum, understanding the EAS itself is still
incomplete and unsatisfactory. The problem is demonstrated
in the following discrepancies between the EAS observation
and the predictions of EAS simulation:
\begin{itemize}
\item
{\bf Number of Muons: } The number of muons measured on the ground,
which is a tracer of hadronic energies in the EAS,
is up to $\sim$100\% larger than that predicted by the standard
air shower simulation (see {\it A.~Yushkov} and
{\it G.~Rodriguez}). It is certain that hadronic interaction
models at UHE, as well as the nuclear effect, is responsible
for this. Incomplete hadronic interaction database
at lower energies may also be contributing to this discrepancy.
\item
{\bf Longitudinal Shower Development: }
Different interaction models give different longitudinal
development and X$_{max}$ distribution. The prediction of $<X_{max}>$, RMS(X$_{max}$)
and their energy development (elongation rate) 
differ among models.
The measured $<X_{max}>$ and RMS(X$_{max}$) are not well
reproduced by the simulations. Cases of good agreement exist,
such as the HiRes/TA measurements and QGSJET/proton simulation,
but this does not sufficiently prove
that ``QGSJET'' is the correct hadronic interaction model at UHE
and the proton is the unique primary ``composition'' of UHECR.
It is interesting to note, however, that existing model predictions
for iron have little differences for $<X_{max}>$, RMS(X$_{max}$) and
its energy development.  
\item
{\bf Lateral Shower Development: } 
The energy of UHECR events measured by the TA/SD is 27\% higher than
the energy measured by TA/FD. The AGASA/SD (plastic scintillator) gives
higher energy than TA/SD, but only by about 9\%.
The systematic uncertainty of
energy measurement is 21\% for TA/FD and $\sim$20\% for TA/SD and AGASA,
meaning that the difference is at the limit of both systematics.
This difference means that the same EM energy measured 600$\sim$800 m away
from the shower core by TA/SD and AGASA/SD is larger than the EM energy
measured by the FD near the shower axis, suggesting the data tends
to have larger lateral development than the simulation predicts.
\end{itemize}

The advent of the LHC is opening up a possibility to
improve this situation drastically.
The LHC maximum beam energy is scheduled to be 7.0 TeV, corresponding
to a 10$^{17}$ eV cosmic ray proton colliding with a proton
target at rest. Hadronic and nuclear interactions at UHEs
can be studied now at accelerators and
the results can be fed back to the hadronic interaction models.
Two very forward experiments at the LHC are particularly relevant.
One is the LHCf measuring the spectrum of neutral particles
($\gamma$, $\pi^\circ$, and neutron) at 0$^\circ$. 
The LHCf is dedicated for cosmic ray related studies.
Another is the TOTEM experiment to measure total and (in-)elastic hadronic
cross sections in the very forward region. It also offers valuable
data to determine the development of EAS in the atmosphere.

In addition to the heavy ion collision data available from the RHIC the
LHC is foreseeing proton-lead collisions next year, which will
contribute to the understanding of the nuclear effect at UHE.
This situation was reviewed and discussed by {\it Y.~Itow}, {\it T.~Sako},  
{\it T.~Pierog}, {\it R.~Engel} and other symposium contributors.
A report of hadronic interaction and simulation WG is also 
available \cite{wg_simulation}.     

Using newly available data from accelerator experiments,
an update of hadronic interaction models
are on the way for QGSJET (from v03 to v04) and EPOS (from v1.99 to -vLHC).
{\it T.Pierog} reported the status of tuning with highlights on the
following observables.
\begin{itemize}
\item 
{\bf Total Cross Section: }
The total inelastic cross section determines the average
depth of the first interaction of UHECRs
in the atmosphere, and affects the speed of shower development
and its fluctuation. The newly available data from TOTEM is being
used for the tuning. 
\item
{\bf Multiplicity: }
This parameter, as well as the inelasticity, affects
the speed of shower development, and the number of muons on the ground.
The data from CMS and ALICE at large angles (small pseudo-rapidity)
are being used for the tuning.
\item
{\bf Inelasticity, or Energy Spectrum at Forward Region: }
The higher the leading particle energy, the slower the shower development
expected. The data from CMS, ALICE and PHENIX are checked with the
predictions of simulations. The p and p${_{\rm T}}$ distribution of $\gamma$'s and
$\pi^\circ$'s obtained by LHCf will be useful for this tuning.  
\item
{\bf (anti-)Nucleon Production at Very Forward Angles: }
The forward-produced nucleon is effective to maintain hadronic
energies in the longitudinal shower development, thus affecting the muon
rate at the ground level. The neutron production
rate being analyzed by the LHCf experiment is eagerly waited.  
\end{itemize}

Auger recently published a measurement of the p-air total cross section using
the X$_{max}$ tail distribution around 10$^{18}$ eV $-$ 10$^{18.5}$ eV 
where the dominant CR composition is considered to be proton \cite{auger_totXsection}. 
In future, with the progress of understanding the CR composition,
a possible appearance of the new phase of hadronic interactions and/or
unknown collective nuclear effects at UHE may be revealed
in the data of EAS development. When it happens, it would become
a good case for the bi-lateral contribution between
the CR physics at UHE and elementary particle physics.

It should be noted that the energy region above 10$^{17}$ eV
is where we expect the galactic to extra-galactic transition of cosmic rays
to take place (see section \ref{sec:anisotropy}). 
On-going low energy extensions of HEAT/AMIGA and TALE
will be soon producing high statistics spectrum and composition data
in this energy range, where the EAS simulation is being calibrated directly
by the precision accelerator data.

\section{Prospects}
\label{sec:futures}
A discussion on the future global facilities and the associated
R\&D were made in the symposium, keeping in mind 
the rapid progress of UHECR research and the still remaining experimental
challenges as described above. 
\begin{itemize}
\item
{\bf Next Generation Ground Observatory: }
Three conceptual designs for the next generation Ground Observatory
with huge aperture (called nGO in this report) were presented by
{\it A.~Letessier-Selvon, P.~Privitera} and {\it S.~Ogio}.
Astrophysical objectives of nGO were briefed
by {\it E.~Parizot}, and interdisciplinary science was discussed
by {\it L.~Wiencke}.
\item
{\bf Space Observatory: }
Observation of UHECRs from space was reviewed by 
{\it T.~Ebisuzaki}, and the mission, performance and
the prototype test of JEM/EUSO were presented by
{\it A.~Santangelo}, {\it M.~Bertaina} and {\it M.~Casolino}.
\item
{\bf R\&D for Radio Detection:}
Radio detection of UHECRs at MHz range
was discussed by {\it O.~Scholten} and
{\it A.~M.~van~den~Berg}. It was
extensively studied by LOPES and CODALEMA, and now by
AERA at the Auger site.
{\it C.~Williams} reported on the GHz radio measurement by the
accelerator beam (MAYBE). Field tests for detecting the air shower
radio signal at the Auger site (AMBER, EASIER and MIDAS), and at KASCADE-Grande
(CROME) were reported by {\it P.~Facal} and {\it R.~Smida}.
{\it J.~Belz} reported a plan of
bistatic radar detection of UHECRs (TARA) at the TA site.
\end{itemize}
\begin{figure}[ht]
  \begin{tabular}{cc}
    \begin{minipage}[b]{0.43\columnwidth}
      \begin{center}
        \resizebox{0.77\columnwidth}{!}{
         \includegraphics{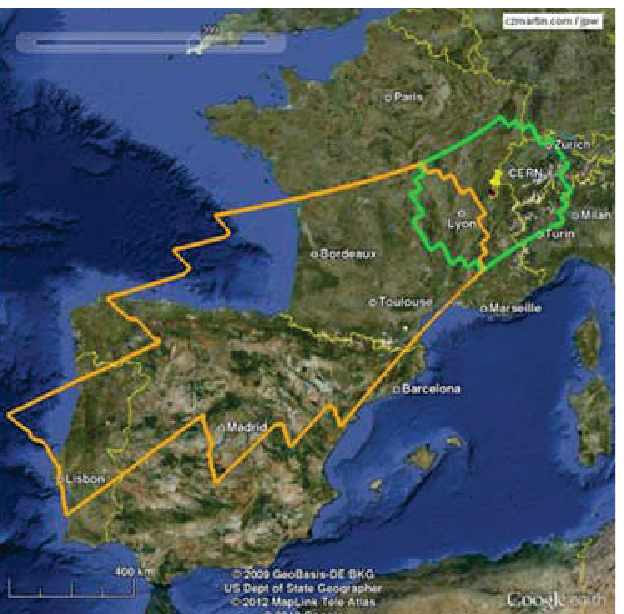} }
         \caption{Footprint of the JEM-EUSO FoV projected on the ground:
         nadir mode (green) and tilted mode (orange) are shown
         \cite{future_space}.}
        \label{fig:area_jem-euso}
      \end{center}
    \end{minipage}
    \hspace{3mm}
    \begin{minipage}[b]{0.53\columnwidth} 
      \begin{center}
        \resizebox{0.96\columnwidth}{!}{
          \includegraphics{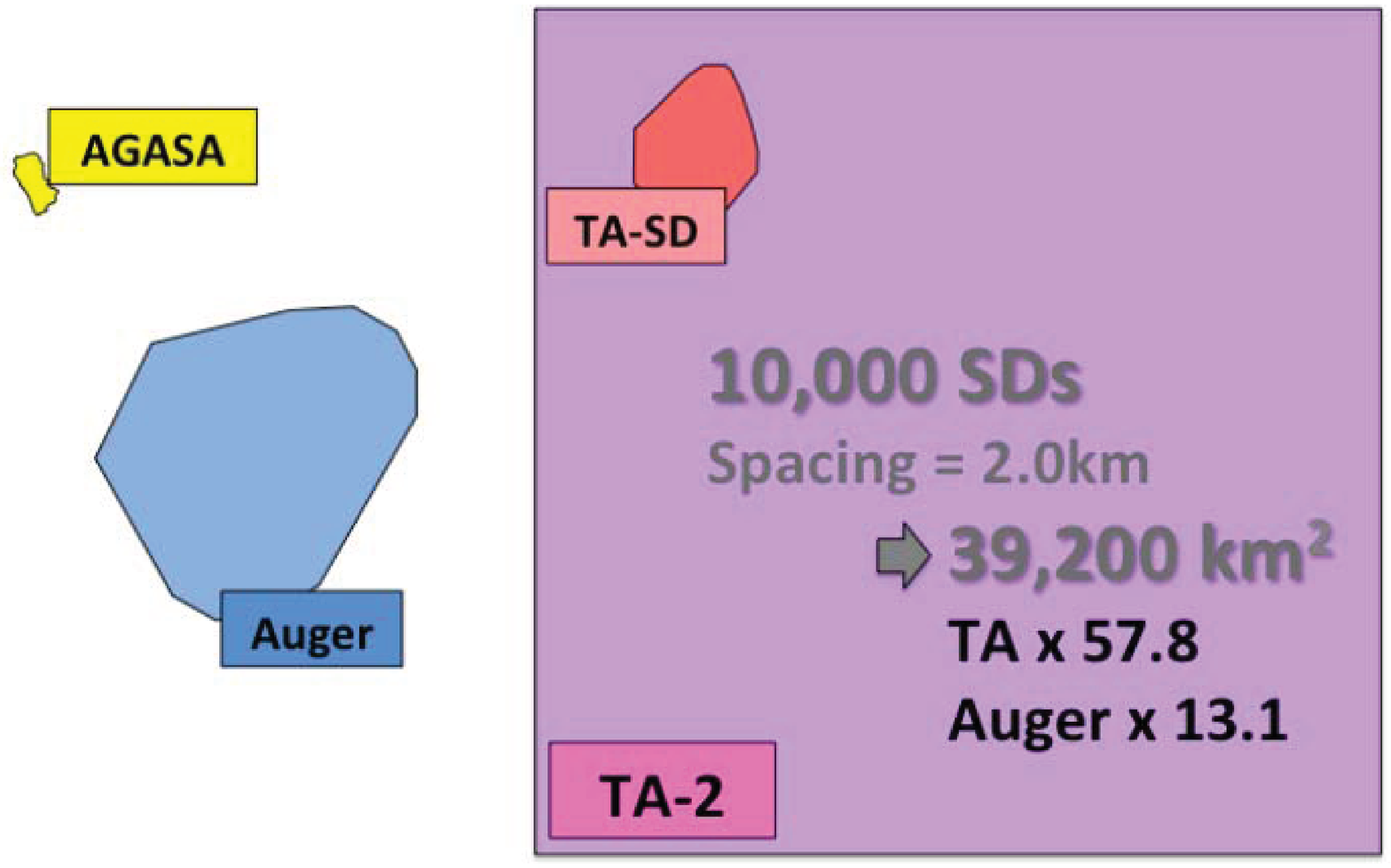} }
          \caption{The covered ground areas of past, present and
          future ground arrays.
          The proposed huge array by S.Ogio \cite{future_emcal} has
          10,000 TA-type plastic scintillator SDs deployed with 2.0 km spacing
          and covers an area of 39.2k km$^2$.}
        \label{fig:area_nGWD}
      \end{center}
    \end{minipage}
  \end{tabular}
\end{figure}
\begin{figure}[h]
  \begin{center}
    \resizebox{0.8\columnwidth}{!}{
      \includegraphics{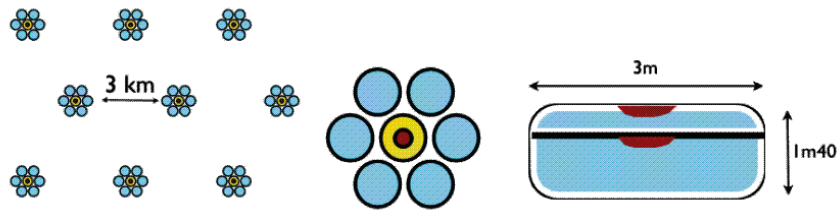} }
      \caption{A possible configuration of the multi-component UHECR Observatory
      proposed by A. Letessier-Selvon \cite{future_muon}. It uses updated
      water tank SDs
      of Auger-type. Vertical segmentation of the tank serves for the EM
      calorimetry (upper thinner layer) and the muon counting (lower part).}
    \label{fig:ngSD}
  \end{center}
\end{figure}

\smallskip\noindent
A majority of the designs and R\&D
presented in the symposium concentrated on realizing a
huge detection area, which is an order of magnitude
larger than that of Auger.
New generation experiments aim to collect $\sim$1,000 events
each year in the trans-GZK energy region (E $>$ 10$^{19.7}$ eV)
to identify the astrophysical origins of UHECRs.
Their features may be compared as follows:

\smallskip\noindent
{\bf Ground vs Space: }
The SDs of Auger and TA measure the UHECRs with 
an angular resolution of $\sim$1.0$^\circ$ and
an energy resolution of $\sim$10\% at 10$^{20}$ eV.
The nGO plans to extend the aperture
maintaining the same level of resolution
for trans-GZK events. This can be achieved 
by tuning the SD detector size and the separation
using the presently available UHECR samples.
A typical aperture of nGO is 55,000 km$^2$ sr
({\it S.~Ogio}).
For Auger and TA, the determination of composition by FD
is statistical and is limited to hybrid events
(duty $<$ 10\%). For nGO, a separate measurement of
the EM energy and the number of muons particularly at each
nGO/SD will give us a handle to
determine the composition event-by-event
({\it A.~Letessier-Selvon}).

The JEM/EUSO aims at
2.5$^\circ$ and 30\% resolution at 10$^{20}$ eV.
The X$_{max}$ resolution is 120 g/cm$^2$ and allows
a separation of $\gamma$'s
and $\nu$'s from  protons and iron. 
The energy threshold is $\sim$10$^{19.5}$ eV.
A large aperture of 100,000 km$^2$ sr (after 20\% duty factor applied)
is obtained by a single telescope looking down on the earth's atmosphere
from the International Space Station (ISS) at a height of 400 km.
A tilt mode will increase the aperture to 350,000 km$^2$ sr
but with a sacrifice in performance. 
The JEM/EUSO design is tuned
to search for astrophysical origins in the GZK horizon,
and the all sky-coverage in the ISS orbit
is advantageous for this purpose.

\smallskip\noindent
{\bf Fluorescence vs Radio: }
A composition measurement by FD continues to be
indispensable in the future nGO. The entire 
nGO area may be covered by an array of wide angle
FD telescopes deployed in a mesh of $\sim$20 km.
Given that the event geometry is reconstructed
using the nGO/SD information, only the time development
of nGO/FD signal is sufficient to reconstruct the EAS longitudinal
development. Such information can be collected
by a small number of PMTs with much larger pixel size than that
of Auger and TA, thus drastically
reducing the cost for construction and maintenance ({\it P.~Privitera}). 
The trigger capability, signal to noise ratio and resolution of X$_{max}$
reconstruction are being studied for this scheme.

When a radio detector is integrated in nGO, it would supply
information of shower longitudinal development 
with high duty ($\sim$100\%), little atmospheric or weather
correction and at low cost. Such a radio detector may be used
in the nGO in place of the now standard FD telescopes.
Many laboratory and field measurements have been
performed to detect MHz, GHz and radio echo signals from EAS. 
The existence of MHz signal is firmly established
by earlier experiments of LOPES, CODALEMA and others,  
and it is known that the signal emission is forward
peaked and has a clear correlation with the geo-magnetic field. 
The GHz signal is expected to have an isotropic and
un-polarized emission from the molecular bremsstrahlung,
but so far the observed GHz signals by CROME and EASIER seem
forward peaked, and not identified as the molecular
bremsstrahlung signal. The forward peak, in both MHz and GHz cases,
means the footprint of the signal on the ground is small
and the radio measurement would be costly, requiring many
antennae to cover the ground.
The TARA radio echo test experiment is now set up on the TA site.
It consists of a 54 MHz, 40 kW television transmitter illuminating
the air above TA, and a set of radio receivers on the far side of TA,
aiming to detect a forward scattered radio signal in coincidence with
the TA measurement.

So far, no proven radio method to replace
the air fluorescence measurement is found, nevertheless,   
the R\&D of radio signal detection continues in order to
understand the nature of signal generation, and with
its application for the X$_{max}$ measurement in mind.
    
\smallskip\noindent
{\bf Tank vs Scintillator: }
The Cherenkov signal of Auger water tanks is larger
for penetrating muons than soft electrons and gammas in the shower,
thus making the Auger/SD more sensitive to a change of
composition with energy.  Muons are also a good measure to investigate
hadronic interactions in the shower development.

The plastic scintillator of TA equally samples the muon and electron,
and the resultant signal is dominated by the outnumbered electrons, or the EM energy.
Since the bulk of the primary energy of UHECRs is transferred to the EM energy,
separate measurement of the EM and muon components in the nGO/SD
would allow us to take both advantages: a good linearity
in the primary energy measurement by the EM component,
and a good sensitivity to the primary composition (and hadronic interaction)
by the muon counting.

For UHE $\gamma$ and $\nu$ detection, a high profile of the water tank
would allow a detection of very inclined EAS down to lower energies
than thin plastic scintillators can achieve. The contrary may
be true for the detection of UHE $\gamma$ ray interactions which contain much fewer 
muons than normal EAS of the same primary energy.

In order to exchange ideas and experiences for the next generation
detector, it was agreed to form a new
working group for future projects.
The JEM/EUSO group also expressed an interest to join
the working group. 
The design of the next generation detector
becomes more sensible and realistic when
the current issues of primary composition and the energy
scale/linearity are understood.
An exchange of analysis method, calibration, simulation,
obtained data and eventually the FD/SD detectors may be
considered among experiments to attack this formidable problem. 
It is encouraging, as a first attempt of many others to be
considered, Auger and TA initiated a photometric and optical
calibration of FDs by flying a common standard light source
in Utah and Malargue by using a GPS controlled octo-copter.
The first flight at TA was completed in October 2012, and the next
flight at Auger site is planned in November.

\section{Summary and Acknowledgement}
\label{sec:summary}
The scene of UHECR is rapidly evolving.
We just settled a long lasting, difficult experimental question
on the energy spectrum of extra galactic cosmic rays;
Yes, there is a ``cutoff'' at 10$^{19.6}$ eV and
``ankle'' at 10$^{18.7}$ eV. The accuracy
of energy measurement is about 20\% now by the new air
fluorescence method, and it is steadily improving.
In the northern hemisphere, a constant
particle composition of proton is measured above
10$^{18.2}$ eV, whereas the composition
is changing from proton to iron in the southern hemisphere.
We eagerly examine our detectors and data analysis to know
whether this difference is the nature or an experimental
artifact. Calibration of air shower simulation is proceeding
with the LHC and other accelerator data.
We do hope it will shed light on the interpretation of X$_{max}$
data. No UHE gammas and neutrinos were detected, and the
top-down acceleration of UHECR is strongly disfavored. Limits of
astrophysical acceleration of UHECR is not seen,
or may have been seen, depending on the composition result
and its model interpretation. 
A transition of cosmic ray sources from galactic to extra-galactic
may be identified in the near future by low energy extensions
of present experiments (HEAT/AMIGA and TALE).
Signatures of UHECR sources come and go, though certain levels 
of anisotropy seem to be an unavoidable consequence of the present
observations. By JEM/EUSO, space may become part of our work field
in the near future. It will seek to identify UHECR origins
in all sky in the trans-GZK region. The R\&D and design of nGO,
a new Ground Observatory, is proceeding. It will accumulate
1,000 events of well observed UHECRs each year
to definitively solve existing
questions on the UHECR, and to explore a new field of physics
in the universe. So, be patient ({\it P.~Privitera}),
and stay tuned.

This summary report is based on the work of UHECR-2012 contributors
and working groups. I have to regretfully admit the introduction
of their works to this report is very incomplete. For more details and
to understand complicated points,
I encourage readers to consult the original papers
in the same proceedings. To make this summary report,
I am largely indebted for
the work of contributors and WGs, but also feel responsible
for my views expressed in this report, and any mistake,
bias and inaccuracy appearing in this report.  The UHECR-2012 was a very active
and enjoyable symposium. High scientific activities and good spirits
of the participants, and the symposium organizers,
are truly appreciated.

\end{document}